\documentclass{aa}  
\usepackage{graphicx}
\usepackage{txfonts}
\usepackage{longtable}
\usepackage{float}
\usepackage{cleveref}
\usepackage{color}
\usepackage{tablefootnote}
\usepackage{multirow}

\DeclareUnicodeCharacter{2212}{-}

\begin{document} 

%


   \title{Physical properties of more than one thousand brightest cluster galaxies detected in the Canada France Hawaii Telescope Legacy Survey}

   \author{A. Chu
          \inst{1}
          \and
          F. Sarron
          \inst{2,3}
          \and
          F. Durret
          \inst{1}
          \and
          I. M\'arquez
          \inst{4}
          }

   \institute{Sorbonne Universit\'e, CNRS, UMR 7095, Institut d’Astrophysique de Paris, 98bis Bd Arago, 75014, Paris, France \\
              \email{aline.chu@iap.fr}
        \and
            Jodrell Bank Centre for Astrophysics, University of Manchester, Oxford Road, Manchester UK
        \and    
            IRAP, Institut de Recherche en Astrophysique et Plan\'etologie, Universit\'e de Toulouse, UPS-OMP, CNRS, CNES, 14 avenue E. Belin, F-31400 Toulouse, France
         \and
             Instituto de Astrof\'isica de Andaluc\'ia, CSIC, Glorieta de la Astronom\'ia s/n, 18008, Granada, Spain}
   \date{}

\titlerunning{Brightest Cluster Galaxies}
%
 
  \abstract
   {Brightest cluster galaxies (BCGs) are very massive elliptical galaxies found at the centers of clusters. Their study gives clues on the formation and evolution of the clusters in which they are embedded.}
   {We analysed here in a homogeneous way the properties of a sample of more than one thousand BCGs in the redshift range $0.15<z<0.7$, based on images from the Canada France Hawaii Telescope Legacy Survey.}
   {Based on the cluster catalogue of 1371 clusters by \citet{Sarron+18} we applied our automatic BCG detection algorithm and identified successfully 70\% of the BCGs in our sample. We analysed their 2D photometric properties with GALFIT. We also compared the position angles of the BCG major axes with those of the overall cluster to which they belong. }
   {We found no evolution of the BCG properties with redshift up to $z=0.7$, in agreement with previous results by \citet{Chu+21}, who analysed an order of magnitude smaller sample, but reaching a redshift $z=1.8$. The Kormendy relation for BCGs is tight and consistent with that of normal elliptical galaxies and BCGs measured by other authors. The position angles of the BCGs and of the cluster to which they belong agree within 30 degrees for 55\% of the objects with well defined position angles.}
   {The study of this very large sample of more than one thousand BCGs shows that they were mainly formed before $z = 0.7$, as we find no significant growth for the luminosities and sizes of central galaxies. We discuss the importance of the intracluster light in the interpretation of these results. We highlight the role of image depth in the modelisation of the luminosity profiles of BCGs, and give evidence for the presence of an inner structure which can only be resolved on deep surveys with limiting apparent magnitude at 80\% completeness m$_{80}$ > 26 mag/arcsec$^2$.
   }


\maketitle
%

\section{Introduction}

Located at the intersection of cosmic filaments in the large scale structures, galaxy clusters present in their center, at the bottom of the cluster potential well, a supermassive galaxy which is also most often the brightest galaxy of the cluster. This galaxy is referred to as the Brightest Cluster Galaxy (BCG hereafter).

It is commonly believed that BCGs are supermassive elliptical galaxies with quenched star formation and little to no gas left. Their gas has been consumed during mergers to form stars at earlier epochs, as  predicted e.g. by \citet{thomas10.1111/j.1365-2966.2010.16427.x}, who find that most of the stellar population present in such galaxies was already formed in-situ before the cluster was formed at z$\simeq 2$. 
However, BCGs with huge reservoirs of molecular gas and strong on-going star formation have been observed and identified \citep[see][]{Mcdonald2016,fogarty2019,castignani2020molecular}. Some of these particular BCGs also have irregular shapes instead of the regular elliptical morphology expected. Sometimes, these can even be reminiscent of jellyfish galaxies, such as RX~J1532+3020, that shows traces of a recent merger, with UV-emitting filaments and knots, indicating recent starbursts \citep[][]{castignani2020molecular}. Such cases remain rare, as shown by \citet{Chu+21}, who reported that only two BCGs out of 98 were blue galaxies (negative rest frame blue minus red color) with a high star formation rate (SFR $>$ 100 M$_\odot$/yr). \citet{Cerulo_2019} found that only 9\% of the BCGs that constitute their sample of 74275 BCGs up to $z$ = 0.35 have colors bluer than 2$\sigma$ from the median color of the red sequence.

BCGs have formed via numerous dynamical and environmental processes such as galactic cannibalism, cooling flows from the central AGN or dynamical friction, but the relative importances of these processes on the growth of BCGs and on the stellar mass assembly are still controversial \citep[see][and references therein]{castignani2020molecular}. Besides, it is unclear if BCGs are still evolving today, as authors find conflicting results.

Observations show that at low and intermediate redshifts (z$<1$), or even at local redshifts, some BCGs can still be undergoing major mergers that could potentially affect the growth of the central galaxy. This agrees with \citet{Bernardi_2009} or \citet{Ascaso_2010}, for example, who find that the sizes of BCGs have grown by a factor of 2 since $z$ = 0.5. 

The galaxies which compose the cluster and constitute the red sequence of the cluster, including the BCG, are mostly red elliptical galaxies with little gas content. Major mergers were shown to have little effect on the mass growth of BCGs and to be unlikely to trigger a new starburst phase \citep[][]{delucia2007}. In fact, \citet{delucia2007} show in simulations that half of the final mass of BCGs is already in place by redshift 0.5. \citet{thomas10.1111/j.1365-2966.2010.16427.x} find that most of the stars in BCGs were already formed before $z$ = 2, and \citet{delucia2007} show that at least 80\% were already formed by $z$ = 3. These studies indicate that it is likely that the stellar population in these galaxies has settled since at least the last 10 Gyrs. This is consistent with \citet{Stott_2011} who find no significant change in the size or shape of these galaxies since $z$ = 1.3. Additionally, \citet{Chu+21} find that the physical properties of BCGs such as their effective radius or surface brightness show little to no evolution since redshift $z$ = 1.8, and conclude that BCGs were thus mainly formed before z~=~1.8. BCGs undergoing major mergers (12 clusters) were also detected in this last study, and were found to have properties that did not differ from those of other BCGs.

Understanding how BCGs were formed and how they evolve can help us to understand how the clusters which host them were formed. BCGs are the result of billions of years of successive galaxy mergers which can leave an imprint on the galaxy. Numerous studies have shown that clusters are preferentially aligned in the cosmic web, along the filaments which connect them, and that neighbouring clusters separated by less than 30~Mpc tend to "point to each other" \citep[see][]{1982A&A...107..338B}. BCGs were also found to share this same tendency with their host clusters: \citet{donahue2015,west2017ten, Durret_2019,  De_Propris_2020, Chu+21} found that BCGs tend to align with the major axis of their host cluster. This means that BCGs tend to have a preferential orientation pointing to the filaments of the cosmic web along which galaxies and groups are falling towards the bottom of the cluster potential well where the BCG is expected to reside. BCGs are aligned by tidal interaction \citep[][]{Faltenbacher2009} and show stronger alignments for brighter galaxies, for rich and more massive clusters, and for small redshifts \citep[see][]{Faltenbacher2009,10.1111/j.1365-2966.2010.16597.x,Hao2011}. \citet{10.1111/j.1365-2966.2010.16597.x} find that the second to fifth ranked galaxies of the cluster also tend to show signs of alignment, although not as strongly as the BCG. \citet{west2017ten} show that other galaxies in the cluster, excluding the BCG, have no preferential orientations in the cluster. 

The present study uses data from the Canada France Hawaii Telescope Legacy Survey (CFHTLS) and concentrates on the redshift range between 0.15 and 0.7. Although the resolution is not as good as that of the HST, we increase the sample size by almost a factor of 20 \citep[there are 74 BCGs in the same redshift range in][]{Chu+21}, making this analysis one of the largest studies on BCGs, thus allowing to go deep into the study of the luminosity profiles of these galaxies.

The intracluster light (ICL) is also brought up in this study. The ICL is composed of stars that were stripped from their host galaxies and are now trapped in the potential of the cluster, but are not gravitationally bound to any individual galaxy in the cluster. This ICL constitutes a very diffuse and faint component of the cluster, 
which is thus very difficult to detect and can merge with the extended envelope of the central galaxy. One challenge is to distinguish the BCG from the ICL. Some works have shown, by studying the velocity dispersion as a function of the distance to the BCG center, an increase in the dispersion observed at longer distances \citep[][]{Cui2014MNRAS.437..816C,Jim_nez_Teja_2021}. This discrepancy shows the border between ICL and BCG. However, spectroscopic observations are, as of now, necessary to achieve this. In the absence of such spectroscopic data, algorithms which attempt to detect the ICL on deep large scale images are being developed and should allow to detect low surface brightness objects even more efficiently on optical images in the near future. This topic is bound to become all the more important with the release of deep sky surveys \citep[see for example][]{Jimenez2016,ellien:hal-03256563}. In this paper, we also aim to discuss and estimate how much the ICL may impact the luminosity profiles of BCGs at these redshifts. 

The paper is organized as follows. The data used as well as the method to detect the BCGs are presented in \Cref{section:catalogue}; we then describe how the luminosity profiles were modeled and analyze the results obtained in \Cref{section:properties}. In \Cref{section:ICL}, we estimate in a preliminary study the impact of the ICL on the models fitted, and in \Cref{section:depth} the impact of the depth of the images. We also measure the alignment of BCGs with their host clusters in \Cref{section:align}. Finally, we discuss the results and present our conclusions in \Cref{section:discussion}. 

Throughout this paper, we assume a $\Lambda$CDM model with H$_{\rm 0}$ = 70 km s$^{−1}$ Mpc$^{−1}$, Ω$_{\rm M}$ = 0.3 and Ω$_{\rm \Lambda}$ = 0.7. We compute the scales and physical distances using the astropy.coordinates package\footnote{https://docs.astropy.org/en/stable/coordinates/}. All magnitudes are given in the AB system.


\section{Obtaining the BCG catalogue}
\label{section:catalogue}

\subsection{The data}

This work is based on the cluster catalogue of 1371 clusters by \citet{Sarron+18}, extracted in the 154~deg$^2$ region covered by the CFHTLS with the AMASCFI cluster finder. The survey is 80\% complete in AB up to a magnitude m$_{\rm i}$ = 24.8 in the CFHTLS i filter, for point sources. The code detects clusters as galaxy overdensities in overlapping photometric redshift slices. Multiple detections that occur in such a configuration are then cleaned using a minimal spanning tree (MST) algorithm. The cluster candidates have a mass $M_{200}>10^{14}$~M$_\odot$ and are limited to redshift $z\leq 0.7$. By running the AMASCFI cluster finder on mock data created using lightcones from the Millennium simulations \citep[][]{Springel05, Henriques12} and modified to mimic CFHTLS data, \citet{Sarron+18} estimated that this cluster sample is $90\%$ pure and $70\%$ complete overall. At $z\leq 0.7$, they find that the purity is fairly constant with redshift $\sim 90\%$, while the completeness steadily decreases with increasing redshift and decreasing cluster mass from $\sim 100\%$ down to $\sim 50\%$ at $z \sim 0.6$ and $M_{200} \sim 10^{14}~M_\odot$. The large number of clusters in this catalogue, with known selection function, allowed \citet{Sarron+18} to discuss the properties and evolution of cluster galaxies with redshift in various mass bins.

Photometric redshifts are available for each galaxy in the field in the CFHTLS TERAPIX T0007 release. These photo-zs were computed with the LePhare code \citep[][]{1999MNRAS.310..540A, 2006A&A...457..841I} based on five filters in the optical.

For the purpose of this work we used the redshift probability distribution function (PDF) for each cluster: ${\rm PDF}_c(z)$, computed by \citet{Adami+20} on the \citet{Sarron+18} catalogue. We note however, that contrary to \citet{Adami+20}, we did not split ${\rm PDF}_c(z)$ with multiple peaks into sub-detections. We made this choice in order to stay as close as possible to the original catalogue released in \citet{Sarron+18}. Briefly, the cluster redshift PDF is computed by summing (stacking) the ${\rm PDF}_{\rm gal}(z)$ of galaxies (provided for each galaxy in the photometric redshift release of the CFHTLS T0007) less than 0.5 Mpc from the cluster center and removing the expected field stacked ${\rm PDF}_{\rm field}(z)$ in that region.


This cluster redshift PDF, ${\rm PDF}_c(z)$, allows us to compute for each galaxy in the cluster vicinity the probability that the galaxy and the cluster are at the same redshift: ${\rm P}_z$. This is done by convolving the redshift PDFs following the formalism of \citet{Castignani16} as in \citet{Adami+20} :

\begin{equation}
P_z \propto  \int {\rm PDF}_c(z)~{\rm PDF}_{\rm gal}(z)~dz
\label{eq:pz}
\end{equation}

\noindent We note that since then \citet{Sarron21} proposed a slightly updated version of this formalism that better accounts for the combined uncertainties of cluster redshift and cluster galaxy redshift in the $P_z$ estimate. However, the correction amounts to a few percent in the worst case and we thus decided to reuse the \citet{Adami+20} results directly in the present work.
By taking into account the distance between the galaxy and the cluster center, we also generated the probability that a galaxy was part of the cluster: ${\rm PDF}_{\rm member}(z)$, following \citet{Adami+20}. 

However, \citet{Sarron+18} detected clusters using galaxy density maps (without magnitude or luminosity weighting) computed in large redshift bins. The exact center assigned to each cluster is then taken as the $S/N$ weighted mean position of its individual sub-detections merged in the MST cleaning process. This implies that the center assigned by \citet{Sarron+18} is close to an unweighted barycentre of galaxies in the cluster region. This position may 
differ significantly from the BCG position in some cases (e.g. in highly substructured clusters). Moreover, tests on simulations showed that the uncertainties on the center coordinates as defined by \citet{Sarron+18} can reach hundreds of kiloparsecs in the worst cases. Considering the distance to the cluster center would thus negatively bias our detection rate.

In the following subsection, we describe our method to detect BCGs on optical images from the CFHTLS. This method makes use of the probability for each galaxy to be at the same redshift as the cluster (P$_z$(z)), and not of the probability to be part of the cluster (${\rm P}_{\rm member}(z)$).

\subsection{Detection of BCGs}

We have retrieved the CFHTLS images from the Canadian Astronomy DATA Centre\footnote{https://www.cadc-ccda.hia-iha.nrc-cnrc.gc.ca/} and identified in each cluster the position of its BCG. The BCG is defined as the brightest galaxy in the cluster that lies within a radius of 1.2 Mpc from the center defined in \citet{Sarron+18}, after filtering out foreground and/or background objects. 

Lack of spectroscopic data has led astronomers to develop methods which only make use of photometric properties. In \citet{Chu+21}, spectroscopic redshifts of the clusters allowed to extract accurately the red sequence, which was then used to identify the BCG. With the present data set, we rely on photometric redshifts to distinguish cluster members from field objects.

Similarly to \citet{Chu+21}, we first proceeded by removing foreground galaxies, taking into account the uncertainties on the redshift. Spectroscopic redshifts, if available, were retrieved from NED to remove all sources which are not within the redshift 68$\%$ confidence interval. Point sources were identified in NED, or via the CLASS\_STAR parameter (CLASS\_STAR < 0.95) in SExtractor. To identify foreground galaxies, we calculated the pseudo absolute magnitude (at the cluster's redshift) for each object. Indeed, foreground galaxies would, in that way, appear abnormally bright ($\rm M_{\rm abs} < -26$). Edge-on spirals were excluded as well by filtering out any object with a major-to-minor axis ratio higher than 2.6.


In order to identify the BCG among the remaining galaxies in the catalogue, we measure the signal to noise ratio (SNR) at the galaxy's coordinates on the density map from \citet{Sarron+18}, and consider the galaxy's probability to be at the cluster's redshift (refer to \Cref{eq:pz}). The SNR, compared to the SNR peak (SNR$_{\rm peak}$) defined in a radius of 5 Mpc centered on the cluster center, gives an information on the location of the galaxy in its host cluster. Taking into account the SNR of the galaxy measured on density maps, instead of simply taking the distance to the cluster center as defined in \Cref{section:catalogue}, allows to consider the size and extent of the cluster as well. Indeed, BCGs were shown not to always lie at the very center of their host cluster \citep[see][and references therein]{Chu+21}, so defining a strict limit in distance appears to be hazardous. In an attempt to define the size of the cluster, we compute the SNR in the background of the density maps, $\langle \rm SNR_{\rm bkg} \rangle$, in a ring between 2 and 3 Mpc from the defined cluster center. All objects with SNR $<$ SNR$_{\rm lim}$, with SNR$_{\rm lim}$ = $\langle \rm SNR_{\rm bkg} \rangle$ + 3$\sigma_{\rm SNR, bkg}$ and $\sigma_{\rm SNR, bkg}$ the SNR RMS of the pixels in the background, are considered not to be bound to the cluster, and are thus rejected. Similarly, to determine if a galaxy is part of the cluster, we compare its probability to be at the cluster's redshift, P$_{\rm z}$, with the same probability computed for galaxies in the background, $\langle \rm P_{\rm z, bkg} \rangle$. The background is again taken in a ring between 2 and 3 Mpc from the cluster center. Objects with P$_{\rm z}$ $\geq$ P$_{\rm z, lim}$, with P$_{\rm z, lim}$ = $\langle \rm P_{\rm z, bkg} \rangle$ + 3$\sigma_{\rm Pz, bkg}$ and $\sigma_{\rm Pz, bkg}$ the P$_{\rm z}$ RMS of the galaxies in the background, are considered as belonging to the cluster, and others are rejected. Any object which doesn't agree with these two conditions is eliminated as a BCG candidate.

These limits are well defined if the cluster has a simple elongated shape, and if the signal related to the cluster is not contaminated by another cluster or filament on the density map. Otherwise, the presence of such structures can increase the noise in the background, resulting in too high dispersions for the background SNR and P$_{\rm z}$. This can result in a limiting SNR$_{\rm lim}$ higher than the SNR at the peak of the density map, or a limiting P$_{\rm z, lim}$ higher than a probability of unity, which renders the detection impossible. 
In such cases, we redefine the SNR and P$_{\rm z}$ lower limits and set SNR$_{\rm lim}$ = SNR$_{\rm peak}$ - 2 and P$_{\rm z, lim}$ = 0.70. These limits were chosen after testing different values, and return the best detection rate for our method. We explain how this rate was estimated in the following.

In order to evaluate our method, detections by the algorithm were validated or corrected individually. Two of us visually inspected every image, and compared the position of the detected BCG to the distribution of galaxies on the density maps. We also confirmed that no brighter galaxy in the catalogue was more likely to be the actual BCG, by comparing their SNR and probability $\rm P_z$ to that of the detected BCG. We consider that a brighter galaxy with $\rm P_z$ or SNR similar to the BCG, but slightly below, and close to the defined limits, is more likely to be the BCG. If one galaxy is considered to be a better candidate, upon our verification, we replaced the BCG detected by our algorithm by the new candidate. We estimate that the method successfully detected about 70\% of the BCGs in our sample. For the remaining 30\%, the BCG assigned by the algorithm is not the best candidate we chose upon inspection, and we thus correct manually the detection in our final catalogue.

From our final catalogue of BCGs, we construct a sub-sample of 496 BCGs with known spectroscopic redshifts. We can thus confirm that the BCGs selected in this sub-sample are indeed the BCGs of their clusters, and better estimate the detection rate of our algorithm. We find that 70\% of these BCGs automatically assigned by the algorithm were accurately detected.

There are 133 clusters, i.e. 10\% of the initial sample of 1371 clusters, which we excluded, as we were still uncertain after verification as to which galaxy was the BCG. These clusters have a SNR $\geq$ 4 on the density maps generated by \citet{Sarron+18}. This is consistent with \citet{Sarron+18}, who estimate their catalogue to be 90\% pure. The missing 10\% BCGs might thus correspond to the 10\% false detections in the initial cluster catalogue.
We compare the distribution in redshift of the 10\% of clusters which have no BCG in our catalogue, and the variation of the cluster catalogue purity with redshift in \citet{Sarron+18}, to check if the 10\% of "false" detections we excluded indeed correspond to the 10\% of false detections from \citet{Sarron+18}. And, indeed, we find similar distributions in redshift. The fraction of clusters in our initial catalogue with no BCG detected increases with redshift, as they compose only $\sim$ 5\% of clusters at $z$ < 0.4 (averaged), and $\sim$ 10\% at $z$ $\geq$ 0.4. The purity in \citet{Sarron+18}, similarly, decreases with redshift. For SNR $\geq$ 4, the catalogue is $\sim$ 95\% pure at $z$ < 0.4, and $\sim$ 90\% pure at $z$ $\geq$ 0.4. We can thus assume that the clusters we excluded are indeed the 10\% impurity from \citet{Sarron+18}.

Consequently, our final catalogue of BCGs is supposed to be nearly perfectly pure. We can not state that it is 100\% pure as we do not have spectroscopic redshifts for all objects and can not certify that the selected galaxy is indeed the BCG. We can only identify which galaxy is the best candidate given the information that we have. Our final sample consists of 1238 detected BCGs and is available at the CDS\footnote{http://simbad.u-strasbg.fr/simbad/}. 

Making use of our sub-sample of clusters with spectroscopic redshifts, we estimate the error on the photometric redshift of our whole sample of CFHTLS detected BCGs. 66\% of these BCGs have their spectroscopic redshift included within $z_{\rm clus} \pm 0.025 \times (1 + z_{\rm clus})$, the expected 1 $\sigma$ uncertainty on the cluster photometric redshift \citep{Sarron+18}. On the other hand, 2.6\% of these galaxies have their spectroscopic redshift away by more than 3 $\sigma$ from the photometric redshift of the cluster, i.e, the absolute difference between the two redshifts is bigger than $3 \times 0.025 \times (1 + z_{\rm clus})$. This validates that the photometric redshift uncertainties of the BCG and the clusters are well defined, and that virtually all the BCGs in our final catalogue are \textit{bona fide.}

Colors of the detected BCGs were also computed to better estimate the fraction of "blue" BCGs in that redshift range. We define the (g-i) color by considering the magnitudes measured in the CFHTLS g and i filters, in a 35 kpc aperture diameter. We apply a K-correction \citep[given by an EZGAL][model for elliptical galaxies, see \cite{Chu+21}]{bruzual2003stellar} and correct for reddening by galactic extinction. We consider a galactic reddening law $\rm A_{\rm V}$ = 3.1, reddening values for the CFHTLS filters from \citet{Schlafly_2011}, and dust maps from \citet{schlegel1998maps}. We consider that a galaxy is blue if its (g-i) color is negative. As a result, we find that there are 7\% blue BCGs (89 BCGs) in our sample. These BCGs with blue colors also tend to be at higher redshifts: 68 out of the 89 blue BCGs (76\%) are at $z$ $>$ 0.5. Their photometric properties will be discussed and compared to those of red BCGs in the following.


\section{The properties of the BCGs}
\label{section:properties}

\subsection{Method}

\begin{figure}[h]
\begin{center}
  \includegraphics[width=9cm]{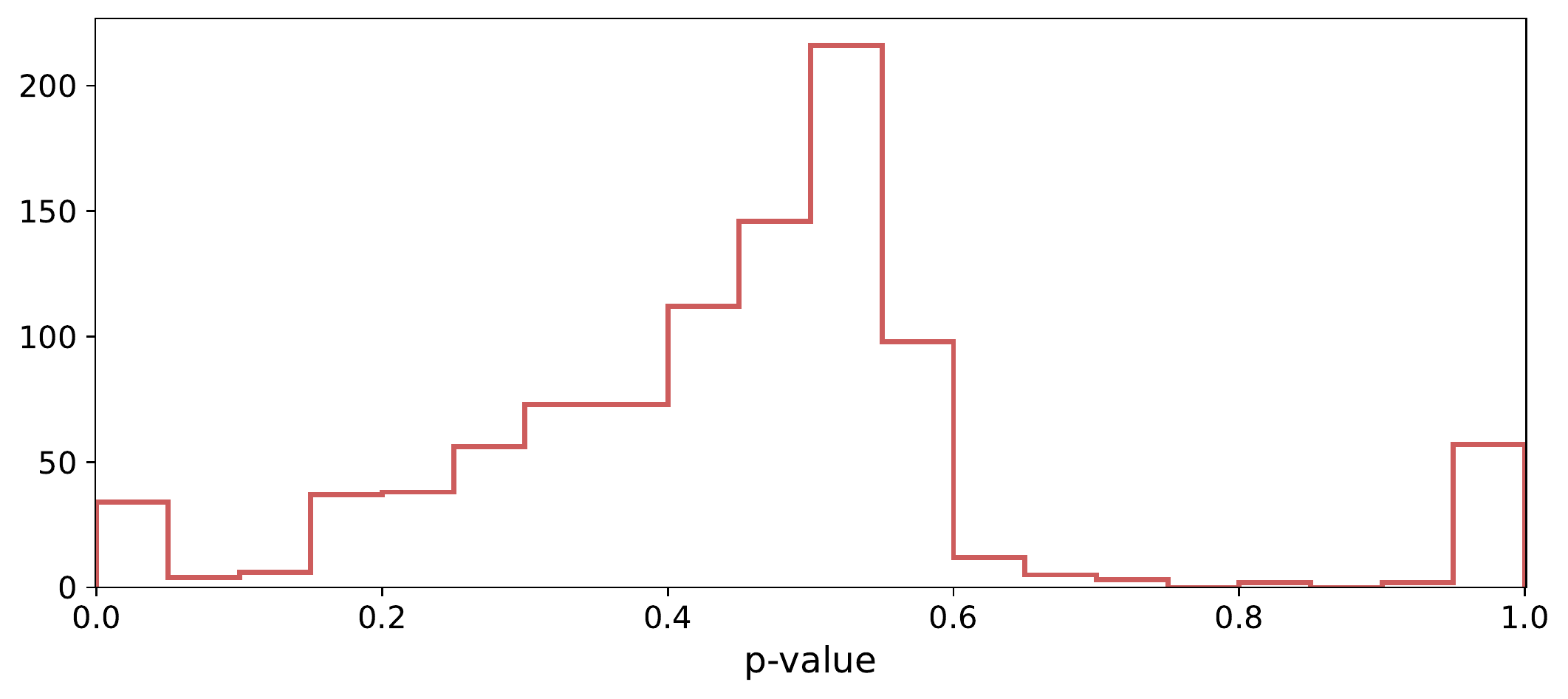}  
\caption{Histogram of the computed p-values. BCGs which could not be fitted with a single component have a default p-value = 0, and those which could not be fitted with two components have a p-value = 1.}
\label{fig:histo_pvalue}
\end{center}
\end{figure}

We fit the 2D luminosity profiles of the 1238 BCGs of our sample with GALFIT in the CFHTLS i band. Up to redshift $z$ = 0.7, this filter falls above the 4000~\AA\ break and hence enables us to consider a homogeneous red and old stellar population for all the BCGs. The method used is the same as described in \citet{Chu+21}: we first mask neighbouring sources using SExtractor segmentation maps and sharp-substracted images \citep[see][]{marquez1999near, marquez2003detection} and model the PSF with PSFex \citep{Bertin11}. Initial parameters are given by SExtractor  by modeling the galaxy with a bulge and a disk. 
We first run GALFIT to fit the BCGs with one S\'ersic component, trying different values of the S\'ersic index $n$ between 0.5 and 10 until the fit converges. 
We reject any non physical fit with effective radius larger than half the size of the fitting region, which is to say $R_e \leq 2.5\ r_{Kron}/2$ pixels, where $\rm r_{\rm Kron}$ is the Kron radius measured by SExtractor. 
Then, we try to add a second S\'ersic component to better model the inner part of the galaxy. If the fit with a single S\'ersic profile converges, we use the output parameters for the single S\'ersic model as initial guesses for the external component of the two-S\'ersic model. Initial parameters for the inner component are taken once again from SExtractor by considering the parameters returned for the bulge component.
If the model with one S\'ersic component did not manage to converge, we reinitialize the parameters for the external component and inner component with SExtractor.

The choice between a one-S\'ersic and two-S\'ersic model is made with a F-test \citep[as was done in][]{margalef2016MNRAS.461.2728M,Chu+21}. The F-test is a statistical test which relies on the residuals and the number of degrees of freedom of two models. The computed p-value, which depends on the F-value (a ratio of reduced $\chi$ squared), must be lower than a probability P$_0$ in order to reject the null hypothesis: if two models give comparable fits, the p-value tends to unity. On the other hand, if the second, more complex model gives significantly better results than the first one, this value tends to zero and we can reject the null hypothesis. We assigned a p-value = 0 to BCGs which could not be fitted with a single component, and a p-value = 1 to those which could not be fitted with two components. Here, P$_{0}$ is defined as the limit between the low p-values computed and the higher values, which is around P$_0=0.15$ (see \Cref{fig:histo_pvalue}). This value of P$_{0}$ is lower than that defined in \citet{margalef2016MNRAS.461.2728M} and \citet{Chu+21}, as the p-value is computed taking into account the number of resolution elements. Since the resolution of the CFHTLS is not as good as that of the HST, the value of P$_{0}$ defined here appears lower. We also checked visually the fits and residuals of galaxies which have p-values close to this limit, to make sure that P$_0$ is well defined. Indeed, we verify that for a p-value p $\geq$ P$_0$ the residuals of the one-S\'ersic model and two-S\'ersic models are similar, and for p $<$ P$_0$, the galaxy profile is better fitted with two components.

\subsection{Results}

\begin{figure}[h]
\begin{center}
  \includegraphics[width=9cm]{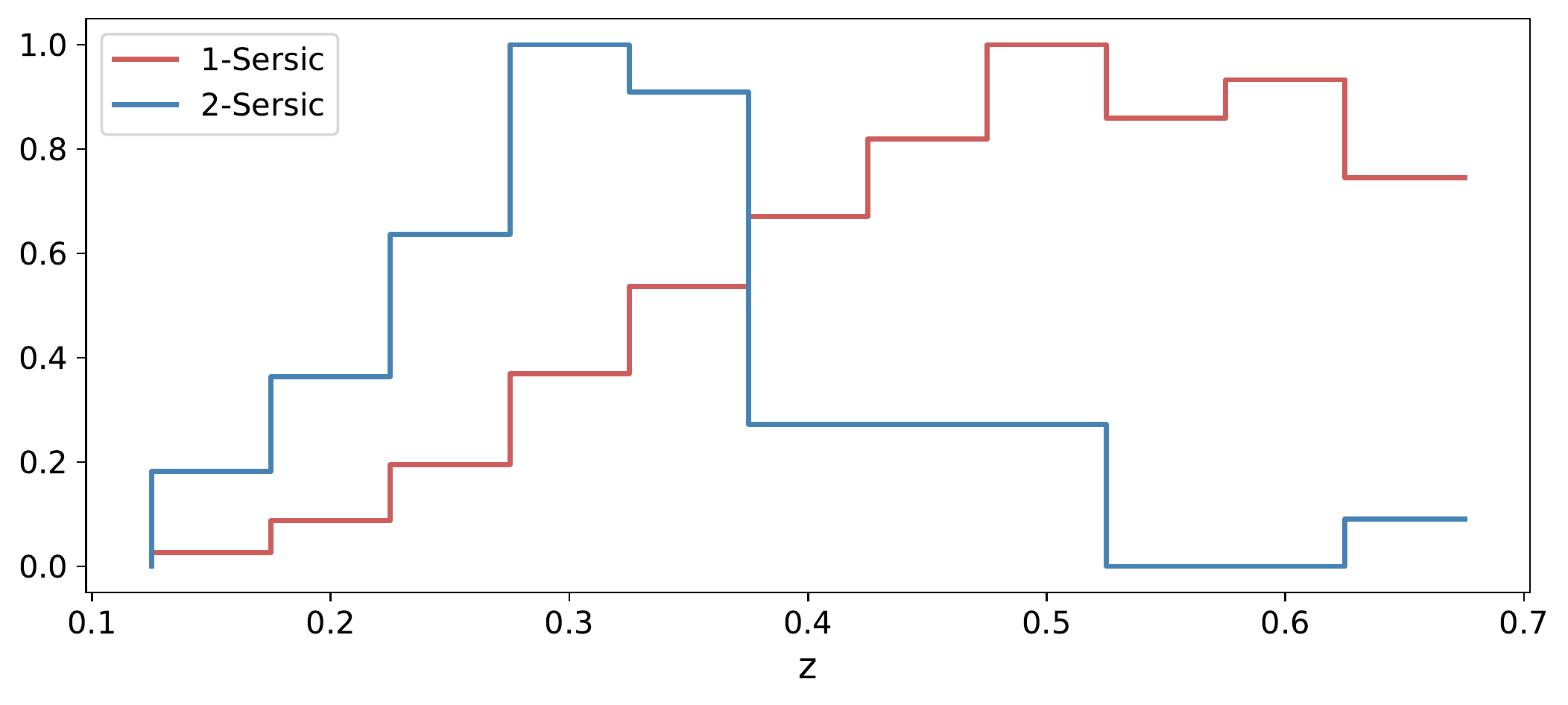}  
\caption{Normalized histograms of the distribution of redshifts $z$ for BCGs fitted with one (red) and two (blue) S\'ersic models.}
\label{fig:histo_z_model}
\end{center}
\end{figure}

\begin{figure}[h]
\begin{center}
  \includegraphics[width=9cm]{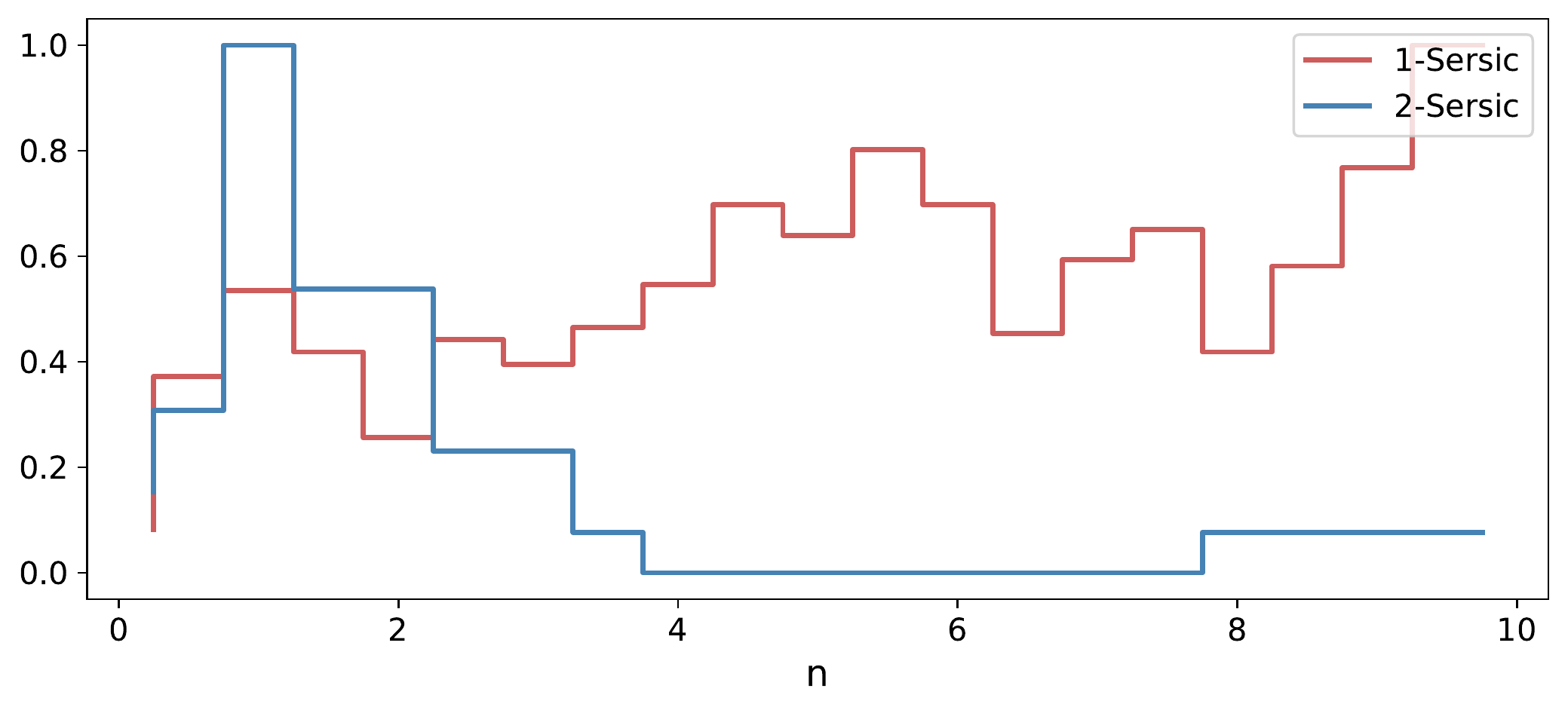}  
\caption{Normalized histograms of the S\'ersic indices n for BCGs fitted with one (red) and two (blue) S\'ersic models. For BCGs fitted with two components, we consider the S\'ersic index of the outer component.}
\label{fig:histo_n}
\end{center}
\end{figure}

Out of the 1238 detected BCGs, 30 were not successfully fitted with either model, bringing our sample to 1208 galaxies. We then only consider galaxies with relative errors on the effective radius, mean surface brightness, and S\'ersic index smaller than 20\%. Excluding the objects with large uncertainties, we end up with 1107 BCGs which are better modeled with:

\begin{itemize}
    \item one-S\'ersic component: 930 BCGs (84\%)
    \item two-S\'ersic components: 177 BCGs (16\%)
\end{itemize}

As in \citet{Chu+21}, we also exclude all galaxies fitted with two S\'ersics that have an inner component (the component with the smaller R$_{e}$) which contributes more than 30\% to the total luminosity of the galaxy. Indeed, in the following, we consider that the outer component of the two-S\'ersic model contains most of the light of the galaxy. This will allow a comparison with a one component S\'ersic model. 

Our final sample is thus made of 974 BCGs with:

\begin{itemize}
    \item one-S\'ersic component: 930 BCGs (95\%)
    \item two-S\'ersic components: 44 BCGs (5\%)
\end{itemize}

Among these, we have 80 blue BCGs (8\%) which were all well modeled with a single S\'ersic component.

In the following, we consider the measured properties of the outer component of the two-S\'ersic BCGs, which will be compared to the properties of the one-S\'ersic BCGs.

We show the normalized redshift distribution of the sample for the two chosen models in \Cref{fig:histo_z_model}. We find one-S\'ersic component BCGs at all redshifts whereas we mainly find BCGs better fitted with two components at lower redshifts (77\% of BCGs better fitted with two S\'ersics are at redshift $z$ $<$ 0.4). 

The histogram of the BCG S\'ersic indices is shown in \Cref{fig:histo_n}. The one-S\'ersic model fits BCGs with all values of the S\'ersic index n, and the distribution appears mostly flat. When a two-S\'ersic model is required, the outer component has a very small value of n (mostly between 1 and 2) and the distribution appears more peaked.

Similar results were found in \citet{Chu+21}. A first natural interpretation could be that, as the redshift increases, the spatial resolution decreases. At lower redshift, galaxies being more resolved, it is possible to distinguish the inner component from the outer component in some galaxies. As a result, galaxies at higher redshifts would be correctly fitted with a single profile, as the center is not correctly resolved. However, \citet{Chu+21} have shown that this is not true: degrading the resolution of low redshift BCGs to the resolution at $z$ = 1 returns comparable results, implying that the models are resolution-independent. Moreover, we find similar distributions of $z$ and n with the HST and CFHTLS samples, which strengthens our point, since the HST resolution is much better than that of the CFHTLS. 


\begin{figure}[h]
    \begin{center}
    \includegraphics[width=9cm]{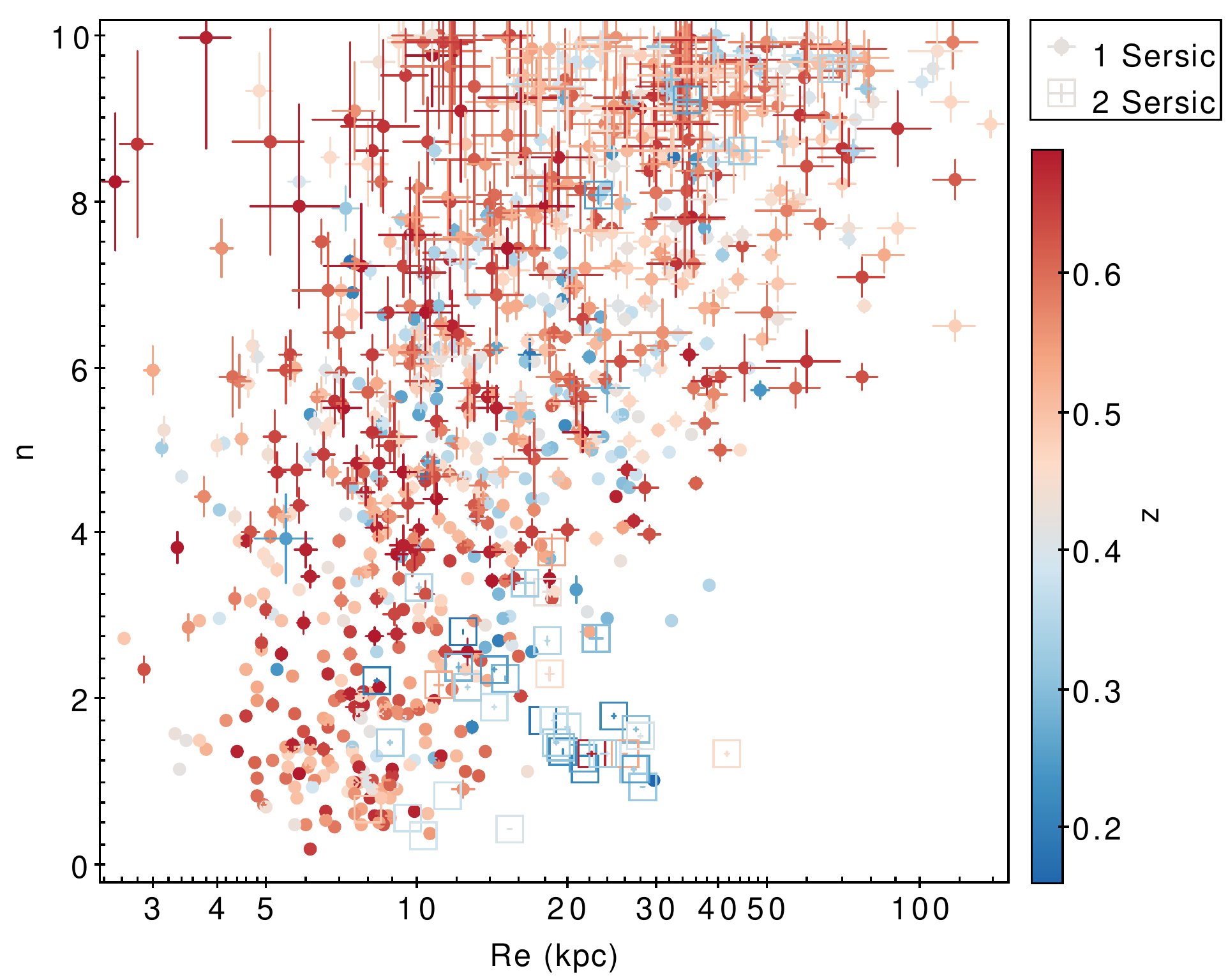}  
    \caption{S\'ersic index $n$ as a function of effective radius $R_e$, color-coded with redshift.
    Dots correspond to BCGs fitted with a single S\'ersic and empty squares to BCGs fitted with two S\'ersics. For BCGs fitted with two components
    we consider the properties of the outer component.}
    \label{fig:Re_n}
    \end{center}
\end{figure}


We display in \Cref{fig:Re_n} the S\'ersic index $n$ as a function of the effective radius $R_e$ in logarithmic scale, color-coded with redshift. Low-redshift objects (z < 0.4) are mainly concentrated in a zone with small S\'ersic index (n < 2) and large effective radius (R$_{e}$ > 20 kpc), and they are also two-component BCGs. When considering only those well fitted with one S\'ersic, we find that the S\'ersic indices increase as a function of the logarithm of the effective radius. Here, we find: 

$n = (5.13 \pm 0.21)\ log(R_{e}) + (-0.29 \pm 0.26)$ (with a correlation coefficient R = 0.62 and significant with a p-value p $<<$ 10$^{-5}$)\footnote{Linear regressions were made using the python scipy lingress function: https://scipy.org/}. In the following, we consider R = 0.20 as the minimum value showing a faint trend, and define p = 0.05 as our significance level.

\begin{figure}[h]
\begin{center}
  \includegraphics[width=9cm]{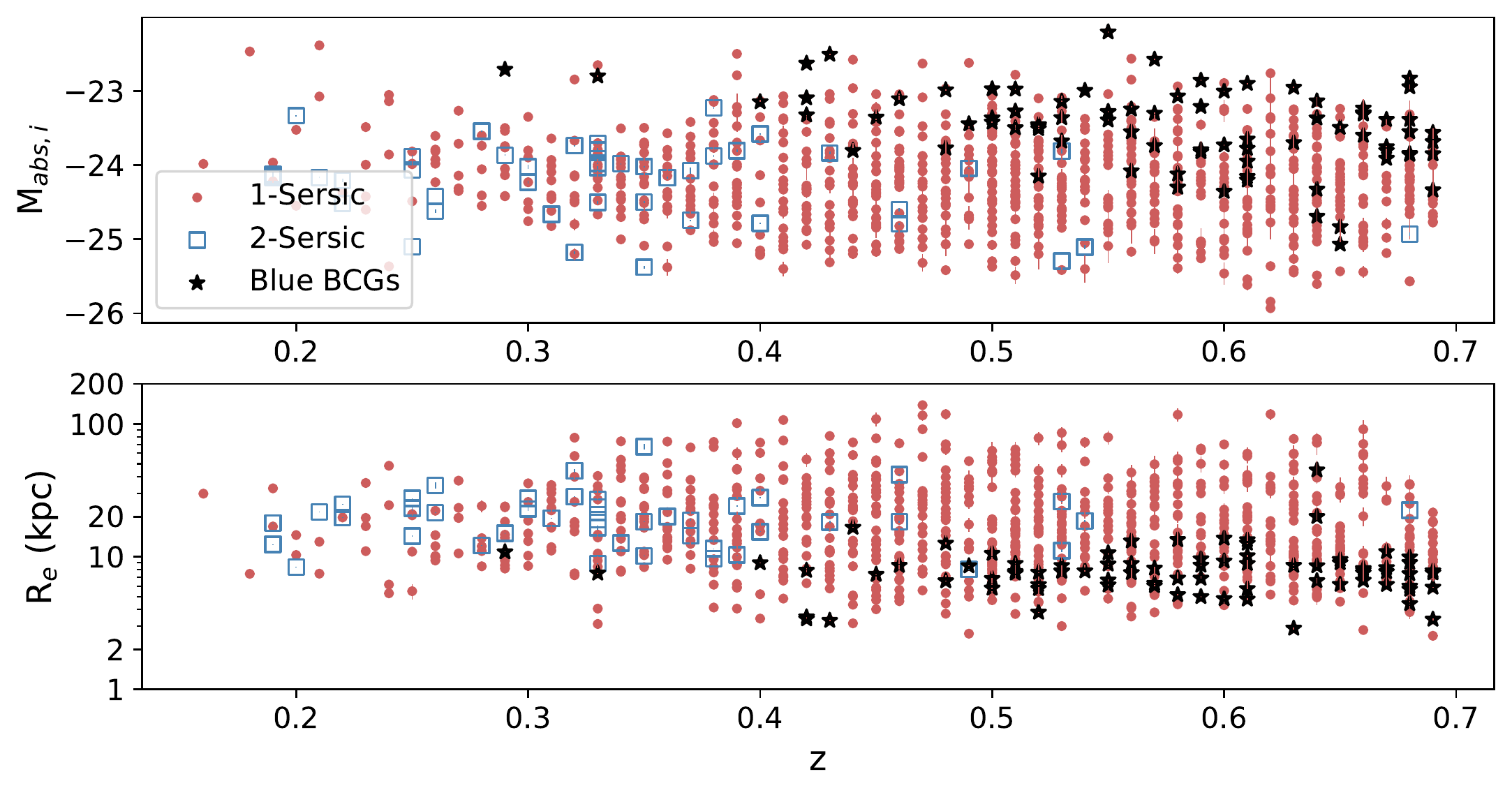}
\caption{Absolute magnitude (top) and effective radius (bottom) measured in the CFHTLS i filter, as a function of redshift. BCGs fit with a single component are represented by red dots, BCGs better fit with two components have their outer parameter represented by empty blue squares. Blue BCGs (see \Cref{section:catalogue}) are identified by dark stars. For BCGs fitted with two components
we consider the properties of the outer component.}
\label{fig:Magabs_Re_z}
\end{center}
\end{figure}

\begin{figure}[h]
\begin{center}
   \includegraphics[width=9cm]{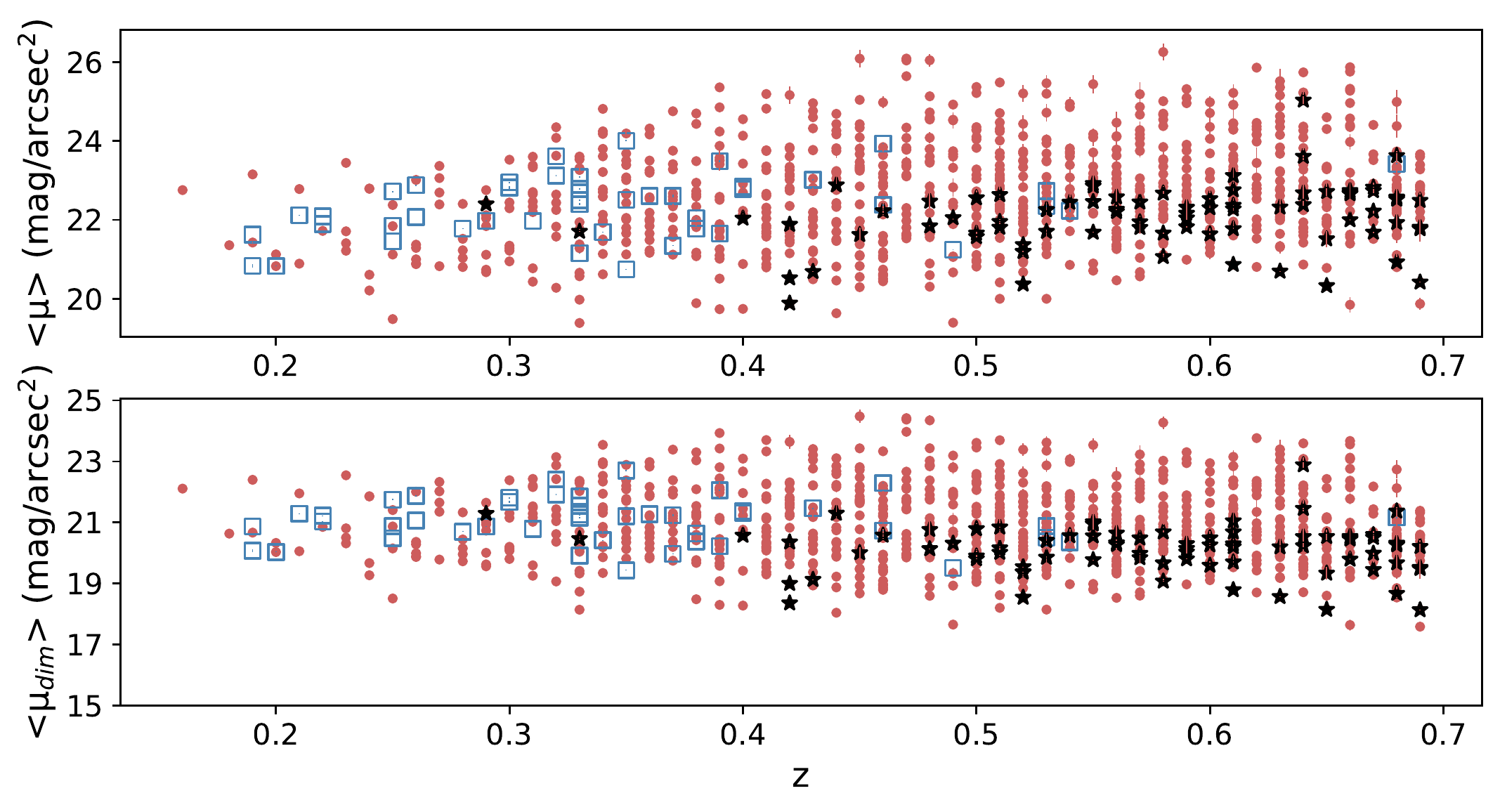}  
\caption{Mean surface brightness as a function of redshift not corrected (top) and corrected (bottom) for cosmological dimming.
The symbols are the same as in \Cref{fig:Magabs_Re_z}. For BCGs fitted with two components we consider the surface brightness of the outer component.}
\label{fig:MeanMu_z}
\end{center}
\end{figure}

The absolute magnitude and effective radius as a function of redshift are displayed in \Cref{fig:Magabs_Re_z}. Absolute magnitudes range between $-26$ and $-22$ with no dependence with redshift. The effective radius is also redshift-independent. 
A very faint trend for BCGs to become brighter and bigger with redshift up to $z$ = 1.8 was observed in \citet{Chu+21} (correlation coefficient R = $−0.29$ with a p-value p = 0.007 for the absolute magnitude to become brighter, and R = $−0.40$ in logarithmic scale with p < 10$^{-3}$ for the size of BCGs to increase). Within the same redshift range as the present study, they measure no correlation for either of these two properties (R = 0.11 for the absolute magnitude, and R = 0.27 for the effective radius, with p = 0.31 and p = 0.01 respectively). By increasing by more than a factor of ten the sample size, we therefore confirm that BCGs have not grown in luminosity or size since $z$ = 0.7 (lower correlations R = 0.06 and R = 0.14 respectively, with p = 0.06 and p $\ll$ 0.05).


The mean surface brightness, not corrected and corrected by a factor of $(1+z)^4$ for cosmological dimming, show no significant dependence with redshift (\Cref{fig:MeanMu_z}). 
As a result, none of the measured parameters (brightness, surface brightness, size, S\'ersic index) is observed to evolve with redshift up to $z$ = 0.7.

Blue BCGs, identified by dark stars on the previous figures, do not show any signs of evolution either. But they seem to occupy preferential locations in these relations. Indeed, blue BCGs tend to be, for the most part, at higher redshifts (76\% at $z$ $>$ 0.5), less bright (mean absolute magnitude for blue BCGs M$_{\rm abs,i, blue}$ = -23.5, for red BCGs M$_{\rm abs,i, red}$ = -24.2) and smaller (mean effective radius for blue BCGs R$_{\rm e, blue}$ = 8 kpc, for red BCGs R$_{\rm e, red}$ = 22 kpc) (see \Cref{fig:Magabs_Re_z}), and to have brighter mean surface brightnesses than their red BCG counterparts (mean surface brightness for blue BCGs $\rm \mu_{\rm blue}$ = 22.1, for red BCGs $\rm \mu_{\rm red}$ = 22.6) (see \Cref{fig:MeanMu_z}).

\subsection{Kormendy relation for BCGs}
\label{subsection:kormendy}

\begin{figure*}[h]
\begin{center}
   \includegraphics[width=18cm]{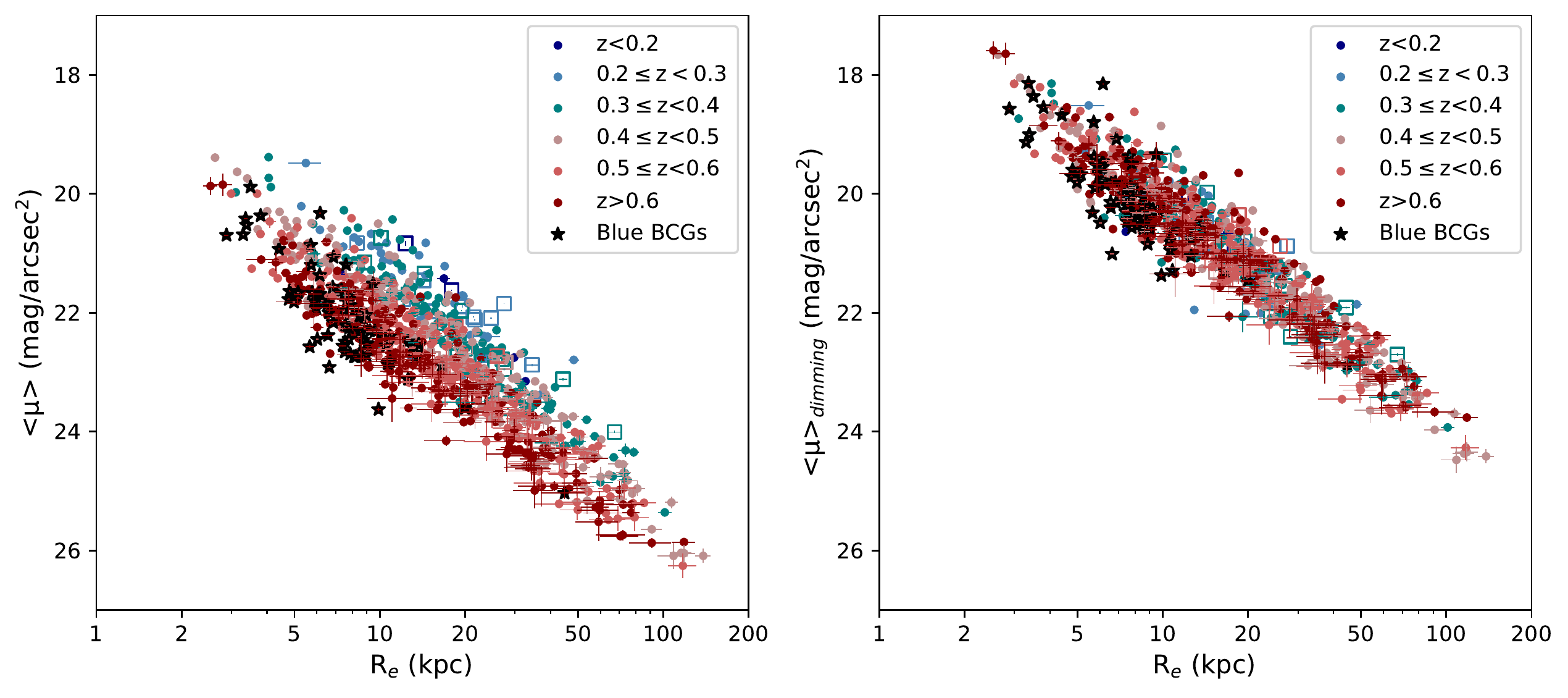}  
\caption{Kormendy relation: mean surface brightness as a function of effective radius before (left) and after (right) correcting for cosmological dimming. The symbols with various colors correspond to different redshift intervals. For BCGs fitted with two components we consider the properties of the outer component.}
\label{fig:Kormendy}
\end{center}
\end{figure*}

The Kormendy relation \citep[][]{1977ApJ...218..333K} between the mean surface brightness and the effective radius is shown in \Cref{fig:Kormendy} before (left), and after (right) correcting for the $(1+z)^4$ cosmological dimming effect. With more than one thousand objets, the Kormendy relation is here very well defined (R > 0.88, p = 0). Without correction for cosmological dimming, we measure the relation:

$\rm <\mu> = (3.34 \pm 0.05) log R_{e} + (18.65 \pm 0.07)$

Similarly to \citet{bai2014inside} and \citet{Chu+21}, we find that the slope of the Kormendy relation stays constant with redshift. 
The relation is also independent of the model used (one or two S\'ersic profiles).

After correcting for cosmological dimming, we find :

$\rm <\mu> = (3.49 \pm 0.04) log R_{e} + (16.72 \pm 0.05)$

This removes the redshift dependence observed on the left figure, and tightens the relation observed. 



\subsection{Properties of the two-S\'ersic BCG inner component}

As in \citet{Chu+21}, we do not find any correlation for any of the properties of the inner component of the two-S\'ersic BCGs with redshift. The similar sample sizes of about 40 BCGs in \citet{Chu+21} and the present paper do not enable us to better constrain the inner part of these galaxies, and we do not have good enough statistics for our analysis to become significant. Still, we can note that, compared to the outer component, the inner component tends to be brighter by at least one magnitude (we remind the selection criterion we applied on two-S\'ersic BCGs to retain only those with an inner component that do not contribute much to the total luminosity of the galaxy) and tends to be smaller by at least a factor of 3 in size.

The Kormendy relation is also very well defined for the inner component at smaller effective radii and brighter mean surface brightnesses (R = 0.93, p $\ll$ 0.05) than the relation illustrated on \Cref{fig:Kormendy}. For the inner component, the relation uncorrected for cosmological dimming is:

$\rm <\mu> = (4.69 \pm 0.29) log R_{e} + (17.94 \pm 0.15)$

\section{Effect of the ICL on luminosity profiles of galaxies}
\label{section:ICL}

Despite its faint nature, the ICL may contribute in the outskirts of BCGs and have an influence on their luminosity profiles at large radii. We try here to quantify how much the ICL affects the BCG profiles in model fitting.

\begin{table}[htbp]
\setcounter{table}{0}
\begin{minipage}{9cm}
\begin{center}
  \caption{Sample of the seven clusters from \citet{Jimenez2018} studied and successfully modeled with GALFIT. 
  The columns are: full cluster name, coordinates of the BCG, and redshift.}
  \label{tab:tab0}
\begin{tabular}{ c |  r  r  l  }   \hline
Name & RA$_{\rm BCG}$ & DEC$_{\rm BCG}$ & ~~z \\ 
 & (J2000) & (J2000) &  \\
\hline                                   
Abell 2744 & 3.59204 & -30.40573 & 0.306  \\
Abell 383 & 42.01412 & -3.53921 & 0.1871  \\
Abell 611 & 120.23672 & 36.05658 & 0.288  \\
MACS J1115.9+0129 & 168.96625 & 1.49862 & 0.349  \\
MACS J1149+2223 & 177.29874 & 22.39854 & 0.5444  \\
RX J2129.6+0005 & 322.41648 & 0.08923 & 0.234  \\
MS 2137-2353 & 325.06316 & -23.66114 & 0.313  \\
\hline         
\end{tabular}
\end{center}
\end{minipage}
\end{table}

We make use of the ICL and background images kindly provided by Y. Jimenez-Teja to estimate the effect of the ICL on the shape and photometry measured by GALFIT. \citet{Jimenez2018} study the ICL fraction in a sample of clusters from the CLASH and Frontier Fields (FF) surveys observed with the HST.
We compared the fits obtained with GALFIT on the original HST images with those obtained after subtracting the ICL using the maps provided by Y. Jimenez-Teja. We prefer to use these HST data rather than our current CFHTLS sample for this study, as we have spectroscopic redshifts available for the HST sample, better spatial resolution, and the clusters studied in \citet{Jimenez2018} have deep images on which the ICL was well studied and detected. 
To check which model between the one-S\'ersic and the two-S\'ersic models fits best our galaxies, BCGs are first modeled with GALFIT on the original images. We find that all BCGs need a second component according to the F-test described in \Cref{section:properties}.

We then subtract the ICL and background from our images and run GALFIT on these final images, which only contain the BCG. 
We then compare the returned parameters, and check if subtracting the ICL allows to remove the inner component needed on the original images. Indeed, this should allow us to understand if the two-S\'ersic model BCGs observed mostly at low redshifts (z $\leq$ 0.4) are an effect of the ICL being more easily detectable at lower z.

From a sample of eleven clusters up to $z$ = 0.54, four BCGs could not be fitted properly with GALFIT with one or two components. The remaining seven clusters studied here are shown in \Cref{tab:tab0}. 
We find that all BCGs, even after subtracting the ICL, are still better fitted using two S\'ersic profiles. 
Thus, we deduce that the ICL does not affect the inner structure of the galaxy, and the need for a second component is not caused by the ICL. By comparing the parameters obtained on the original and ICL subtracted images, by letting all parameters free in GALFIT, we found that the presence of the ICL could disturb the profile of the outer component of the model (in particular, higher effective radius by about a factor of two or three). 
To better estimate how much the ICL can affect our models, we choose to model one more time the BCGs on the original images, but we fix the inner component with the parameters obtained on the images without ICL. Indeed, as the ICL mainly affects the outskirts of the profile, the inner region of the BCG is supposed to be hardly modified. By fixing the inner component, we make sure that we only take into account the differences caused by the ICL.

\begin{table*}[h]
\setcounter{table}{1}
\begin{minipage}{1.\textwidth}
\begin{center}
  \caption{Parameters obtained from fitting the luminosity profiles with two S\'ersic components. The columns are: full cluster name, absolute magnitude, mean effective surface brightness, effective radius, S\'ersic index, for the outer component (left) and inner component (right). For each cluster, the parameters for the outer component are given for the original images (top row) and the ICL subtracted images (bottom row) with fixed inner component.}
  \label{tab:tab2}
\begin{tabular}{ c  | c c c c | c  c c c }   \hline
 &  \multicolumn{4}{c}{External component} & \multicolumn{4}{c}{Inner component}  \\

Name & m$_{\rm ABS}$ & $<\mu_{\rm e}>$ & R$_{\rm e}$ & n & m$_{\rm ABS}$ & $<\mu_{\rm e}>$ & R$_{\rm e}$ & n \\ 
 & (mag) & (mag/arcsec$^{2}$) & (kpc) &  & (mag) & (mag/arcsec$^{2}$) & (kpc) &  \\
\hline                                   

\multirow{2}{*}{Abell 2744}               & -25.93		& 24.76		 & 155.33	 & 3.13 & \multirow{2}{*}{-23.52}	      & \multirow{2}{*}{19.39}		   & \multirow{2}{*}{4.34}	   & \multirow{2}{*}{2.47}  \\
  			  & -24.38		& 21.96		 & 21.04	 & 0.81 & 		      & 		   & 		   & 	   \\
\hline
\multirow{2}{*}{Abell 383}                & -25.20		& 22.01		 & 38.09	 & 0.83 & \multirow{2}{*}{-24.53}	      & \multirow{2}{*}{19.83}		   & \multirow{2}{*}{10.22}	   & \multirow{2}{*}{1.96}  \\
  			  & -25.00		& 21.58		 & 28.52	 & 0.63 & 		      & 		   & 		   & 	   \\
\hline
\multirow{2}{*}{Abell 611}                & -25.46 		& 22.01		 & 36.34	 & 1.26 & \multirow{2}{*}{-23.70}	      & \multirow{2}{*}{20.21}		   & \multirow{2}{*}{7.05}	   & \multirow{2}{*}{2.68}  \\
  			  & -25.19 		& 21.61		 & 26.73	 & 1.03 & 		      & 		   & 		   & 	   \\
\hline
\multirow{2}{*}{MACS J1115.9+0129}                & -25.26		& 22.95		 & 46.50	 & 1.68 & \multirow{2}{*}{-22.66}	      & \multirow{2}{*}{19.50}		   & \multirow{2}{*}{2.88}	   & \multirow{2}{*}{1.69}  \\
  			  & -24.74  	    	& 22.22		 & 26.21	 & 1.06 & 		      & 		   & 		   & 	   \\

\hline
\multirow{2}{*}{MACS J1149.5+2223}                & -26.94		& 25.72		 & 276.22	 & 3.85 & \multirow{2}{*}{-22.81}	      & \multirow{2}{*}{20.78}		   & \multirow{2}{*}{4.23}	   & \multirow{2}{*}{1.88} \\
  			  & -25.24		& 23.04		 & 36.75	 & 1.23 & 		      & 		   & 		   & 	   \\
\hline
\multirow{2}{*}{RX J2129.7+0005}                 & -25.54              & 22.18		 & 44.48	 & 2.02 & \multirow{2}{*}{-22.09}	      & \multirow{2}{*}{19.15}		   & \multirow{2}{*}{2.24}	   & \multirow{2}{*}{1.22}  \\
  			  & -25.27              & 21.75		 & 32.14	 & 1.63 & 		      & 		   & 		   & 	   \\
\hline
\multirow{2}{*}{MS 2137-2353}                  & -25.11		& 20.71		 & 16.30	 & 2.1  & \multirow{2}{*}{-22.21}	      & \multirow{2}{*}{20.95}		   & \multirow{2}{*}{4.80}	   & \multirow{2}{*}{3.21}  \\
  			  & -24.86		& 20.34		 & 12.27	 & 1.69 & 		      & 		   & 		   & 	   \\

\hline         
\end{tabular}
\end{center}
\end{minipage}
\end{table*}

The parameters measured with two components can be found in \Cref{tab:tab2}.
For all seven BCGs, the absolute magnitude M$_{ABS}$ of the external component is brighter after removing the ICL (with a difference $\Delta$M$_{ABS}$ $\leq$ 2). After subtracting ICL, BCGs also have brighter effective surface brightnesses, with Abell~2744 presenting the biggest difference of almost 3 mag/arcsec$^{2}$. Additionally, for all cases, the effective radii increase in the presence of ICL, for some of them drastically. The measured effective radius can be up to thirteen times bigger when the ICL is still present on the images. This is the case for MACS J1149+2223.

\begin{figure}[ht]
\begin{center}
   \includegraphics[width=9cm]{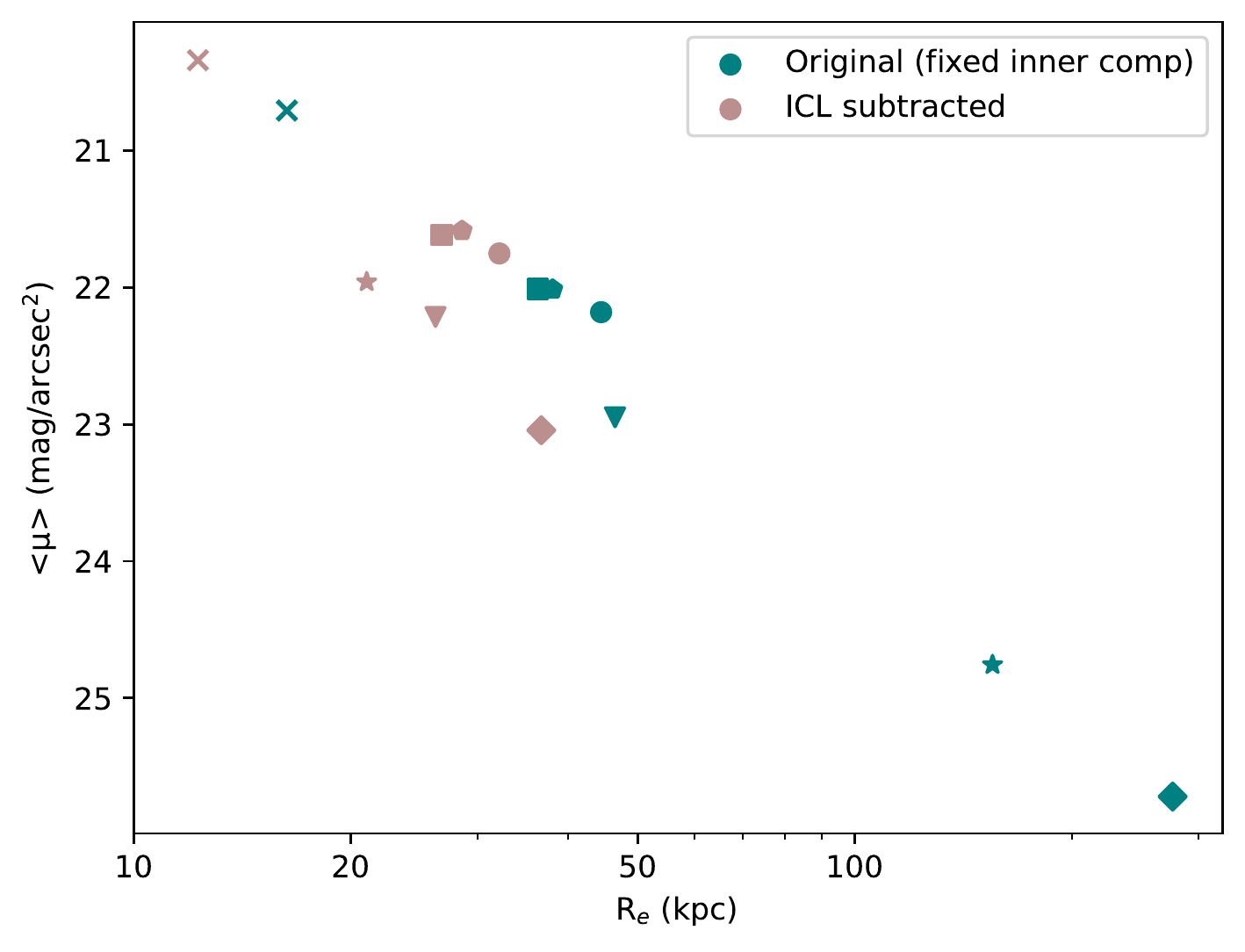}
\caption{Kormendy relation obtained for the 7 BCGs in our sample before (blue) and after (red) subtracting the ICL on the original images. Each cluster is represented by a different symbol.}
\label{fig:kormendy_icl}
\end{center}
\end{figure}

To further illustrate this, we plot the Kormendy relation obtained with the seven BCGs, before and after subtracting the ICL (\Cref{fig:kormendy_icl}). The relation after subtracting the ICL is shifted at lower R$_{e}$ and brighter <$\mu$>, which is consistent with our previous remarks. 
The slope of the Kormendy relation does not depend on the presence or not of the ICL.


The outer component profile on ICL subtracted images still presents a low S\'ersic index with n < 2 for all BCGs. The S\'ersic indices without ICL are smaller than the ones measured with ICL, resulting in a flatter profile in the outskirts. 
This is to be expected, as the ICL would extend the profile at higher radii with very faint surface brightness, and the stars which constitute the ICL would blend with the stars that are bound to the BCG in the outskirts. The galaxy would thus appear less compact, bigger, and more diffuse in the presence of ICL. 

Since the component with very low S\'ersic index (n < 2) is still present even after removing the ICL from the original images, we thus conclude that the dichotomy observed in the distribution of S\'ersic indices and redshifts, following the best model used, is not related to the ICL. Drawing any conclusions regarding the evolution of the size of BCGs with redshift can however be tricky, as the ICL can affect the profile at large radius.

We tried to take the ICL into account by adding a third S\'ersic profile when fitting the original images, and by fixing the parameters of the first two components to those obtained on the ICL subtracted images.
Although a S\'ersic profile might not be the best choice to model the ICL, our goal was only to check if GALFIT would be able to detect a third component on top of the BCG. If successful, then we could try to fit three S\'ersic profiles instead of two to our whole sample, in order not to be affected by the ICL contribution.

The test was done on the cluster RX J2129 (chosen arbitrarily among the BCGs which were well fitted previously). 
A third component was successfully detected and was modeled with faint surface brightness ($\mu_{\rm ICL}$ = 25.24), large effective radius (R$_{\rm e, ICL}$ = 190 kpc, i.e. more than four times bigger than that of the outer component) and very low S\'ersic index (n = 0.4). Though this result was expected, as the ICL is by nature a very extended envelope with faint surface brightness, GALFIT returns a very elongated component (b/a = 0.2) whereas the ICL map appears close to circular. We thus conclude that GALFIT does not manage to correctly model the ICL and has difficulties fitting properly components with very faint surface brightness.  Adding a third component, beside being even more time consuming, is not possible with GALFIT to correct for the effect of the ICL on the outer profile of BCGs.

\section{Effect of the depth of the images}
\label{section:depth}

As demonstrated above, the presence of BCGs with two S\'ersic components observed mostly at low redshifts (z < 0.4) is neither due to the lower resolution at higher redshifts, nor to the presence of ICL at low redshifts. Indeed, in \citet{Chu+21}, we degraded the resolution of low redshift clusters to that at redshift $z$ = 1, and in \Cref{section:ICL} we remove the ICL from our images, to check if these two parameters can influence the choice of the best fitting model, according to the F-test. In both cases, we found that we still need two components to properly fit the BCGs which were better modeled with two S\'ersics on the original images.

Although we increase by more than a factor of ten the sample size up to $z$ = 0.7 from \citet{Chu+21}, we find similar numbers for galaxies better fitted by two S\'ersic profiles. Either this is related to the evolution of BCGs, or another observational bias comes into play. To confirm this, we study how the depth of the images affects the model distribution shown in \Cref{fig:histo_z_model} and in \citet{Chu+21}.

\begin{figure}[h]
\begin{center}
   \includegraphics[width=9cm]{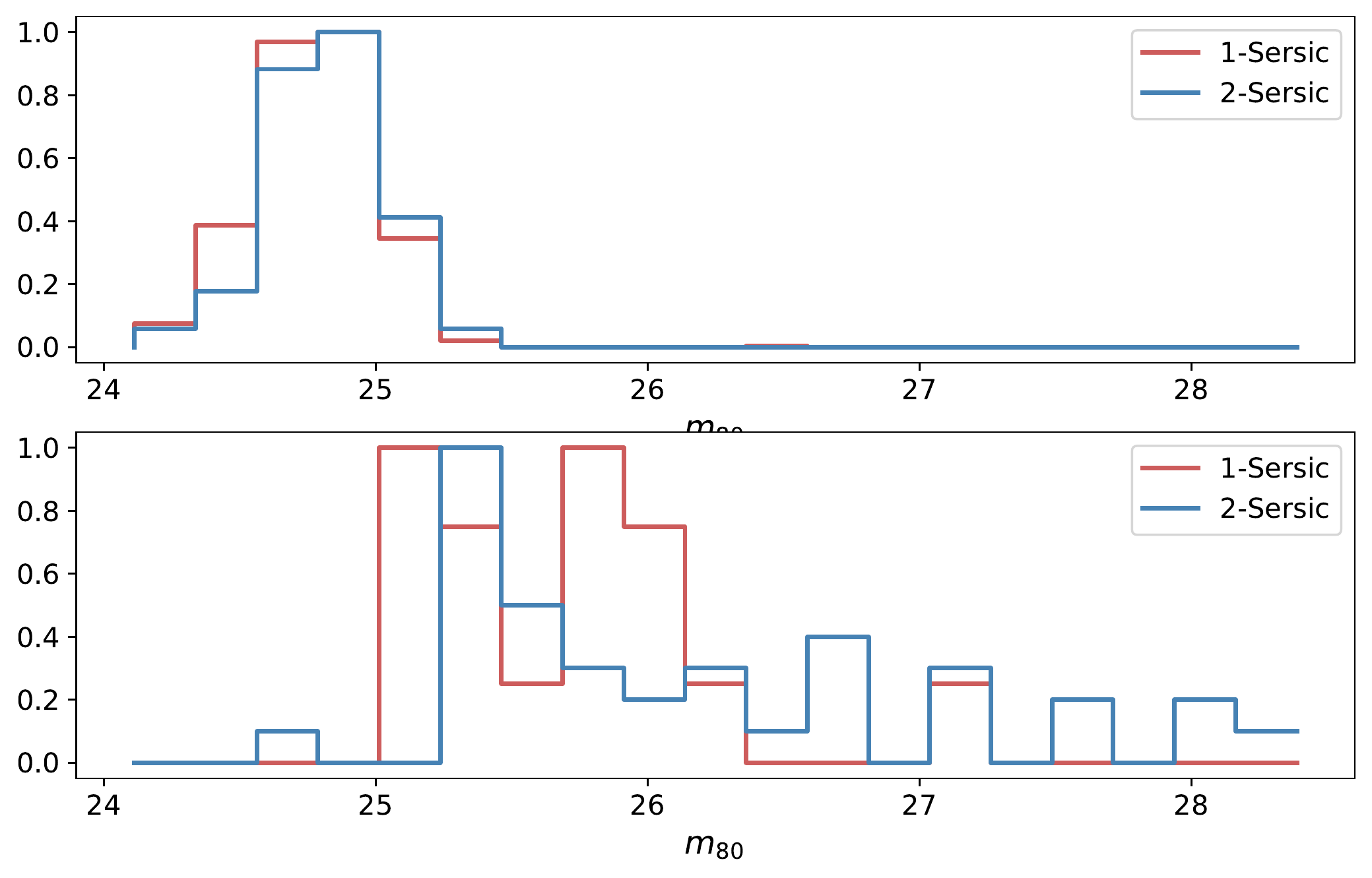}
\caption{Normalized distributions of the magnitudes at 80\% completeness for the sample in this paper (top) and in \citet{Chu+21} (bottom). The red histograms represent BCGs well fitted with a single component, blue histograms are BCGs better fitted with two components.}
\label{fig:hist_model}
\end{center}
\end{figure}

We measure the magnitude at 80\% completeness, m$_{\rm 80}$, of our catalogues and show the distribution of the model chosen as a function of m$_{\rm 80}$ (see \Cref{fig:hist_model}). The images in \citet{Chu+21} obtained with HST are deeper than those used in the present study, based on the CFHTLS. Indeed, in \citet{Chu+21} (\Cref{fig:hist_model}, bottom plot), the distribution of m$_{\rm 80}$ has a peak at m$_{80}$ = 25.4, and can reach m$_{80}$ = 28.0. With our CFHTLS sample, we measure a peak at m$_{80}$ = 24.7 (\Cref{fig:hist_model}, top plot), and no cluster has m$_{\rm 80}$ > 25.5. This is consistent with the value indicated in the TERAPIX documentation\footnote{https://www.cfht.hawaii.edu/Science/CFHTLS/T0007/}. 

With HST data, out of 54 BCGs up to $z$ = 0.7, 37 were better modeled with two S\'ersics, or 69\%. BCGs better modeled with two components have images that go deeper than m$_{80}$ = 25.2. For the deepest images (m$_{80}$ > 26.5), the BCGs which need an additional component become dominant (only one BCG was well fitted with a single S\'ersic, the other 13 BCGs in that redshift range were better fitted with the more complex model). In the present study, in the same redshift range, we find that only 5\% of BCGs need an additional component. Not only the depth of the CFHTLS images does not go as deep as the HST images, but it is limited to a magnitude m$_{80}$ = 25, which is below the magnitude of the peak measured for HST for those fitted with two components.
We can guess that if deeper images were available for the CFHTLS, the structure of the BCGs would be better resolved and the number of 2-component BCGs would increase.

In another attempt to highlight the influence of the depth of the images on the model used to fit the BCGs, we make use of the Deep fields of the CFHTLS. Eight BCGs in our sample were observed both in the Wide and Deep surveys, allowing us to compare directly the effect of the depth of the images on the luminosity profiles of the BCGs. Using the Deep survey, we find that six out of eight BCGs have a two-component structure. On the other hand, all but one of the same objects observed in the Wide survey lack an inner component.
Additionally, the p-values computed on the Wide images (p$_{\rm W}$ > 0.35 on average) are much higher than those obtained on the Deep images (p$_{\rm D}$ < 0.1), indicating that the residuals of the two models tend to become similar as the images become shallower. For one BCG still lacking an inner component on the Deep image, the Wide p-value (p$_{\rm W}$ = 0.59) drops to p$_{\rm D}$ = 0.18 for the Deep image. This suggests that an even deeper image would allow us to resolve the inner component that is currently "hidden". This BCG is also the farthest galaxy in our sample of eight BCGs, at $z$ = 0.65. On the contrary, the BCG which was better modeled with two components on both images is the one at the lowest redshift, $z$ = 0.19. 

Moreover, it may be important to note that, in \citet{Chu+21}, two-S\'ersic BCGs are observed at redshift higher
than $z$ > 0.4 too. Indeed, although they are more dominant at lower redshift and not so much at higher redshift, we still account for two component galaxies up to $z$ = 1.8. By plotting a $z-m_{80}$ diagram, we distinguish two populations at $z$ > 0.7: first, the dominant population of one-S\'ersic BCGs at $z$ > 0.7, modeled on images with m$_{80}$ < 26 mag/arcsec$^{2}$; but also, two-S\'ersic BCGs, modeled on images with m$_{80}$ > 27 mag/arcsec$^{2}$. Consequently, if deeper surveys were available, we could assume that not only would the fraction of two-component BCGs increase at low redshifts, but this would happen at higher redshifts too.

\begin{figure}[h]
\begin{center}
  \includegraphics[width=9cm]{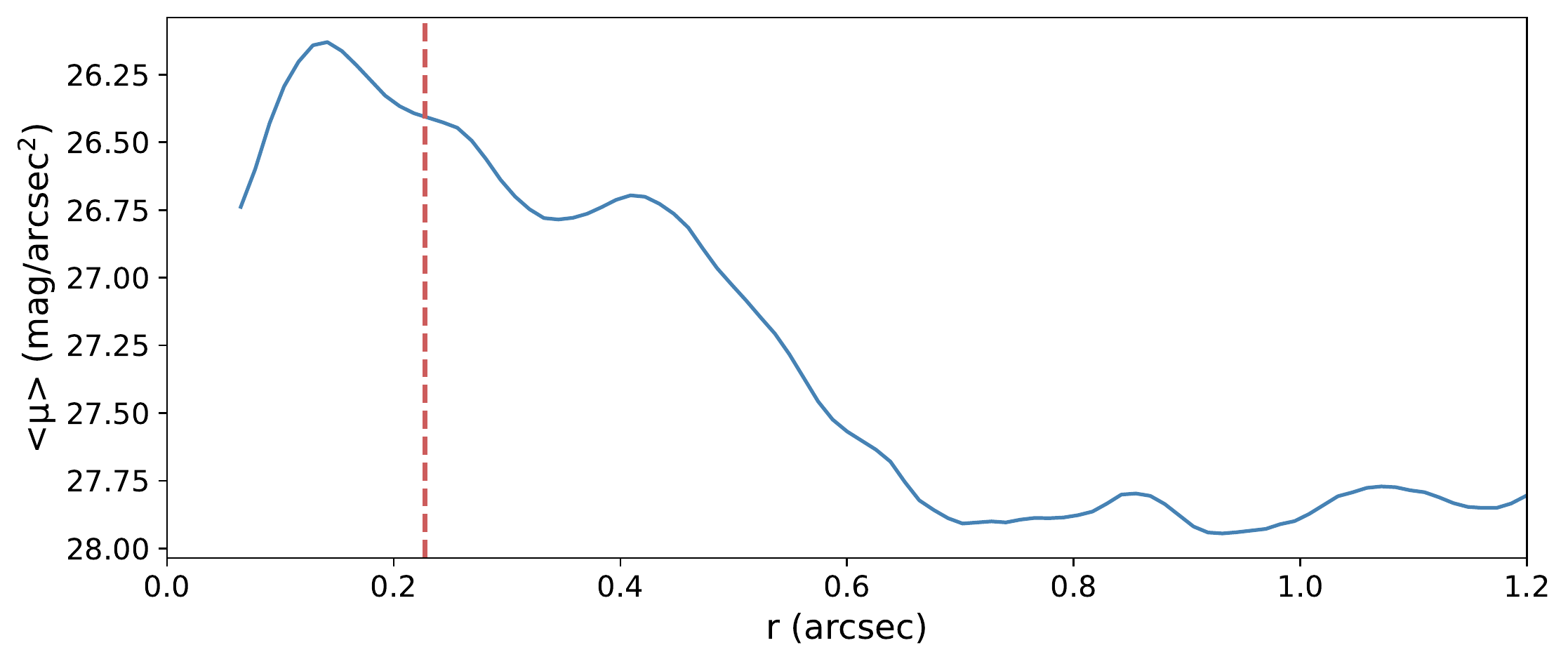}  
\caption{Surface brightness profile of the ICL in the cluster RX J2129. The surface brightness was computed in a circular aperture, centered on the cluster center, with a radius r. The red vertical line represents the Kron radius of the BCG.}
\label{fig:ICL_mu}
\end{center}
\end{figure}

We also check if the ICL can be detected on our images, and thus if the ICL affects our study, by measuring its surface brightness on the images provided by \citet{Jimenez2018} and comparing it with the limit computed for the CFHTLS images. We are able to do this, as images provided by \citet{Jimenez2018} are very deep images limited to magnitude 27.7 in the F814W ACS filter.
Taking the example of RX J2129, the profile in surface brightness is shown in \Cref{fig:ICL_mu}. We find a maximal surface brightness of 26 mag/arcsec$^{2}$ in the center of the cluster. The profile becomes dimmer the farther you go from the center, and reaches a surface brightness of 28 mag/arcsec$^{2}$ at distances r > 0.7 arcsec (r > 2.6 kpc) from the center. When compared to the surface brightness limit of our CFHTLS sample ($\mu_{\rm CFHTLS, 80}$ = 21.5 mag/arcsec$^{2}$), the ICL is too faint to be observed in our images. Indeed, we measure a surface brightness $\mu_{\rm ICL}$ = 26 mag/arcsec$^{2}$ for the ICL, which is fainter that the limit measured in surface brightness on the CFHTLS images. We can thus confirm that the results shown previously are physical and are not affected by ICL.

\section{Alignment of BCGs with their host clusters}
\label{section:align}

In the purpose of studying the alignment of BCGs with their host clusters, we measure their positions angles (PA) and ellipticities with GALFIT.
Out of the 974 BCGs which were fitted by either one or two S\'ersic profiles, 126 have a minor-to-major axis ratio b/a $\geq$ 0.9. As in \citet{Chu+21}, we exclude these galaxies, as an ellipticity close to unity leads to high uncertainties on the measurement of the PA, and so to an ill-defined PA. We thus end up with 848 BCGs. 

We measured the cluster ellipticities by fitting ellipses on the density maps with a 3$\sigma$ clipping, applying the ellipse function in the python photutils package\footnote{https://photutils.readthedocs.io/en/stable/isophote.html}. All clusters with b/a $\geq$ 0.9 were excluded. This brings us down to 639 clusters.

Similarly to \citet{west2017ten}, we measured the PA of clusters by computing the moments of inertia on the galaxy density maps provided by \citet{Sarron+18}, and estimated the uncertainties with bootstrap resamplings. For each cluster, we generated a hundred bootstraps of half of the pixels of the density maps, in a radius R$_{500,c}$ corresponding to the radius within which the mean density is 500 times the critical density of the Universe at the redshift of the cluster, $\rho_{\rm c}$(z). R$_{500,c}$ was computed from the R$_{200,c}$ radius obtained for each cluster in \citet{Sarron+18}. The conversion was done using the relation given in \citet{Sun2009ApJ...693.1142S}: R$_{500,c}$ = 0.669 R$_{200,c}$. This R$_{200,c}$ was derived from the $M_{200}$ estimate of \citet{Sarron+18} inferred from an X-ray derived mass to optical richness scaling relation: 
\begin{equation}
    R_{200,c} = \left( \frac{3 M_{200,c}}{4 \pi 200 \rho_{c}(z)} \right) ^ {\frac{1}{3}}
\end{equation}

We initially wanted to exclude all clusters which present big uncertainties on their PAs. However, we find that most clusters tend to have very high uncertainties of around 45 degrees. This can be explained by the fact that, as stated in \Cref{section:catalogue}, clusters are detected on density maps which cover a wide redshift bin (on average, the width of the redshift bin is $\delta$z = 0.15), which means one detection can overlap with another cluster at a nearby redshift. The presence of filaments, which link clusters in the cosmic web, can also bias the measured PA. For all these reasons, the uncertainties computed by bootstrap resampling can be large if the cluster is not rich (and thus has a low SNR on density maps), if it is circular, or if it is not isolated.

\begin{figure}[h]
\begin{center}
   \includegraphics[width=9cm]{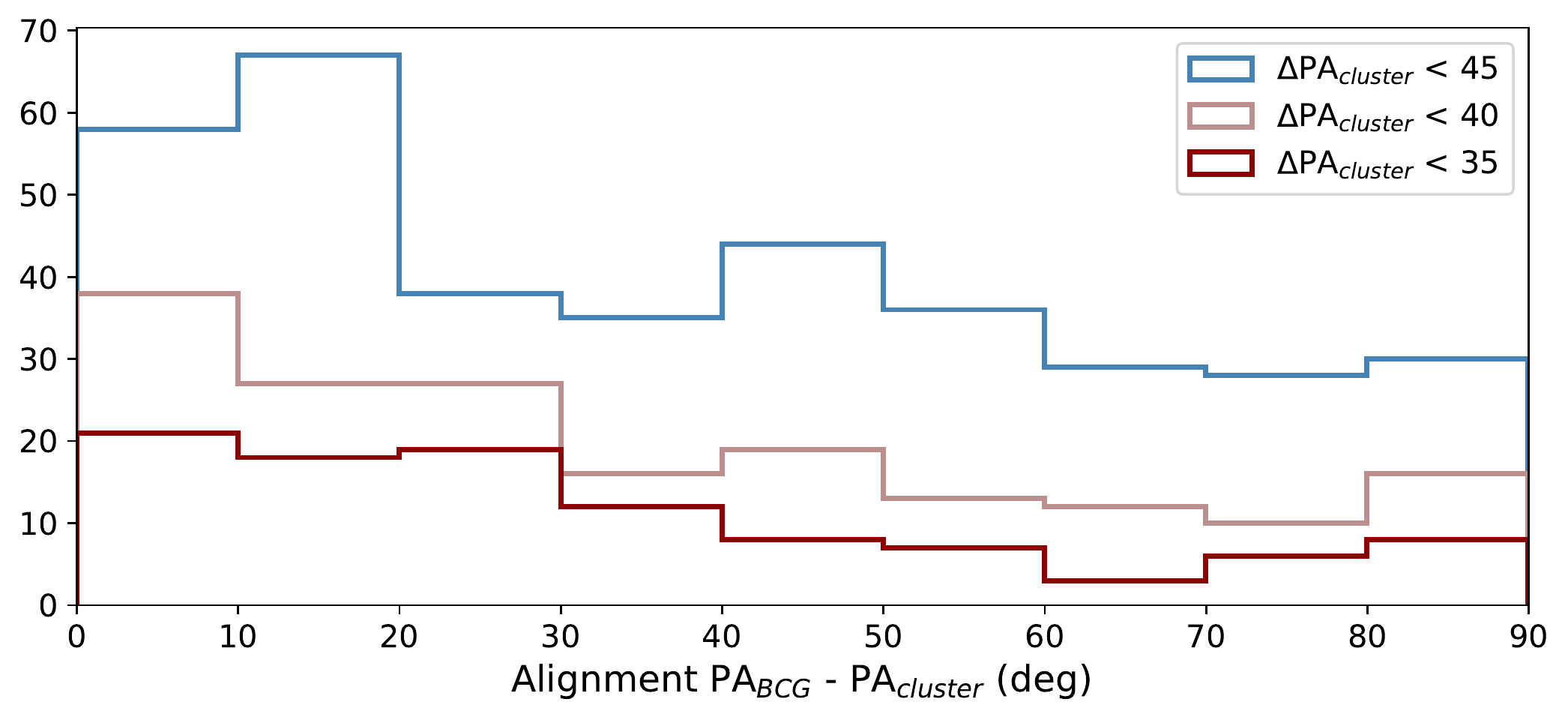}
\caption{Histograms of the alignments (absolute difference of PAs) between BCGs and their host clusters. The histograms were obtained after excluding circular objects and systems, as well as clusters with large PA uncertainties ($\Delta$PA$_{\rm cluster}$ > 45 degrees in blue, $\Delta$PA$_{\rm cluster}$ > 40 degrees in light red, $\Delta$PA$_{\rm cluster}$ > 35 degrees in dark red).}
\label{fig:hist_align}
\end{center}
\end{figure}

We chose to cut the samples by removing clusters with uncertainties bigger than 45, 40, and 35 degrees. This brings us down to samples of 420, 203, and 116 clusters and BCGs respectively.

The alignments (differences between the cluster and BCG PAs) for the final samples are illustrated in \Cref{fig:hist_align}. For all three histograms, even with the biggest uncertainties on the PA of the cluster, we still observe a peak at lower differences.
We measure, respectively (for errors of 45, 40, and 35 degrees), 44 $\pm$ 2\%, 51 $\pm$ 3\%, and 57 $\pm$ 4\% of BCGs aligned within 30 degrees with their host clusters (uncertainties on the alignment fractions were computed by bootstrap resampling). On the contrary, only 24\%, 22\%, and 18\% of BCGs differ by more than 60 degrees from the major axis of the cluster. This shows a tendency for BCGs to align with the major axis of their host clusters that will be discussed in \Cref{section:discussion}. To assess the confidence that our observation is not due to random fluctuations in a sample with a finite number of clusters, we computed the expected uncertainty on $f_{\rm random}$ through bootstrap realisations of sampling from a random distribution for $N = 420$ and 116 clusters respectively. This allows us to estimate that the observed alignment is not due to random fluctuations at $3.4\sigma$ ($\Delta PA_{\rm cluster} < 45~{\rm deg}$) and $4\sigma$ ($\Delta PA_{\rm cluster} < 35~{\rm deg}$) respectively.

Furthermore, we find that BCGs in very massive clusters of M$_{\rm cluster}$ > 5 10$^{14}$ M$_{\odot}$ (therefore, the most massive BCGs, by converting cluster mass to BCG mass using for example the relation given in \citet{bai2014inside}), and bigger BCGs with R$_{\rm e, BCG}$ > 30 kpc, tend to be better aligned than the less massive ones. Indeed, all BCGs in that size and mass range, from \citet{Chu+21} (two BCGs) and in the present paper (12 BCGs), are found to be better aligned than 30 degrees with the major axis of their host clusters. It is however difficult to confirm this with less massive galaxies, because of the large scatter in the M$_{\rm cluster}$ vs. |PA$_{\rm cluster}$ - PA$_{\rm BCG}$| (similarly, M$_{\rm BCG}$ vs. |PA$_{\rm cluster}$ - PA$_{\rm BCG}$|) and R$_{\rm e, BCG}$ vs. |PA$_{\rm cluster}$ - PA$_{\rm BCG}$| relations.


\section{Discussion and conclusions}
\label{section:discussion}

Making use of the galaxy cluster catalogue of \citet{Sarron+18}, we developed an algorithm to detect BCGs on optical images from the CFHTLS. We estimate that 70\% of the BCGs in our sample were successfully detected. The final sample of BCGs built and studied in this paper consists in 1238 BCGs.

This method requires large images in order to properly estimate the background (at least 2 Mpc to be far enough from the cluster centre), as well as images in several filters, to obtain a good photometric redshift estimate of the galaxies in the cluster field. With the CFHTLS, five filters were available to fit the objects with the LePhare code, enabling \citet{Sarron+18} to estimate photometric redshifts with a typical accuracy of $0.05\times(1+z)$.

Several studies such as \citet{Mcdonald2016,Cerulo_2019,fogarty2019,castignani2020molecular,Chu+21} have identified BCGs with unusual blue colors, showing signs of recent starbursts. However, such BCGs are scarce, and increasing their statistics is necessary to better understand which processes lead these galaxies to behave differently from their "red" counterparts. By computing the g-i colors of the BCGs, we also estimate the fraction of blue (g-i $<$ 0) BCGs in the Universe up to $z$ = 0.7. We find that 9\% of BCGs in our sample are blue, which is consistent with the estimates given in the cited papers.

By applying the same method as in \citet{Chu+21}, we modeled the luminosity profiles of the BCGs by fitting two models: one with a single S\'ersic component, and one with two S\'ersic components. The model was chosen using the statistical F-test: we observe two populations with a separation at $z = 0.4$, below which some BCGs tend to require an additional component to take into account the brighter bulge. 
Up to $z$ = 0.7, in \citet{Chu+21}, we found that 77\% of BCGs were better modeled with two components, while here only 5\% are better modeled with two S\'ersics in the same redshift range. Even though we significantly increase the size of our sample, the number of 2-component BCGs did not increase, and we find that the fraction of two-S\'ersic BCGs actually decreases.
In order to understand and explain why these galaxies with a more important bulge exist mostly at lower redshifts, we check for any observational bias that could affect our study.


Although the spatial resolutions of this sample and of the sample in \citet{Chu+21} are different, we still find similar distributions for the best model, with two-S\'ersic BCGs  found mainly at lower redshifts (z < 0.4). We also find similar distributions for the S\'ersic indices, with two-component BCGs having lower indices (n < 2), and single-component BCGs presenting indices which cover a wide range between 0 and 10, with a flatter distribution. This hints that the resolution of the images does not play a large role in model fitting. We confirmed this in \citet{Chu+21} by degrading the resolution of HST images at low $z$ and verifying that the fits returned by GALFIT were unchanged.

We already take into account the distance at which the galaxies are observed, by considering a filter that is above the 4000 \AA\ break, and thus, by always modeling the same red stellar population. However, it is all the more difficult to detect objects with faint surface brightness at high redshifts, without deep exposure times. 
We thus pondered if ICL, which has very faint surface brightness, might be detectable at lower redshifts, and would thus constitute the second component observed. 

By removing the ICL contribution, using images provided by \citet{Jimenez2018}, we show that the presence of ICL can affect the outskirts of the galaxy, and can flatten the profiles at large radii. Indeed, the presence of ICL can decrease the outer S\'ersic index, and increase the size of the BCG. Nonetheless, a second component with low S\'ersic index is still needed even after subtracting the ICL. The presence of two-S\'ersic component BCGs with low index at low redshift is thus also not an effect of the ICL. Moreover, the images studied here are not deep enough to detect the ICL, which is too faint, with a surface brightness $\mu_{\rm ICL}$ $\geq$ 26 mag/arcsec$^2$. We conclude that our study is not affected by the presence of ICL.


Lastly, we compare the completeness of our catalogues. We remind that in \citet{Chu+21} 85\% of BCGs at redshift $z$ $\leq$ 0.4 are two-component galaxies, whereas they only represent 16\% of our sample at $z$ $\leq$ 0.4 here. We find that two-S\'ersic component BCGs in \citet{Chu+21} tend to appear on images which have a depth of the order of m$_{80}$ $>$ 26.5. Our current CFHTLS data does not go as deep as the HST images, and the structure of the BCG is thus not as well determined. We could assume that repeating this study with deeper images would reveal the existence of an inner component at $z$ $\leq$ 0.4 for most of the BCGs that are well modeled with a single S\'ersic component in this paper. The presence of such an inner structure would indicate that bulges of BCGs may have formed first, and the extended envelope would have formed later on, at $z$ $\leq$ 0.4. As \citet{10.1093/mnras/stz2706} state, the cores and inner regions of BCGs were already formed long ago and stopped evolving, whereas the outer regions as well as the ICL started developing recently via minor mergers. This would also agree with the assumption that the ICL was formed at later times ($z < 1.0$), as stated by \citet{Jimenez2018} and references therein. \citet{Montes2017} claim that the ICL saw most of its formation happen at $z = 0.5$; this would agree with the discrepancy between the two models we observe around $z = 0.4$, which could hint at a more important contribution of the ICL to the luminosity profile of BCGs at $z$ $\leq$ 0.4.


In fact, in \citet{Chu+21}, two-S\'ersic BCGs are observed up to $z$ = 1.8, even though they are not dominant at higher redshifts. 
These two component galaxies are found on the deepest images with limiting magnitude m$_{80}$ > 27 mag/arcsec$^{2}$. The remaining population of one-S\'ersic BCGs is modeled on images with m$_{80}$ < 26 mag/arcsec$^{2}$. 
The separation between the two models at higher redshifts that depends on the depth of the images highlights the importance of deep surveys. We would expect to detect two components in all BCGs, but this requires deep images and long exposure times.
Even though the depth of the images could be a solution to resolving the structure of these central galaxies, we can still wonder if other cluster properties may be linked to the properties of this inner component. We do not find any correlation between the properties of the inner component and redshift, or with the cluster properties. Moreover, the low sample size of two-S\'ersic BCGs does not enable us to draw any significant conclusions. Deeper surveys are needed to confirm our results and assumptions, and determine any link between the presence of an inner component and BCG growth.

In order to understand if BCGs are still growing today, we looked for correlations between redshift and the physical properties of BCGs measured with GALFIT. We find no evolution as a function of redshift for the effective radius and absolute magnitude. 
In \citet{Chu+21}, no correlation could be found for the mean surface brightness when no dimming correction was applied, up to $z$ = 1.8. However, a trend could be seen up to $z$ = 0.7 (R = 0.29, p = 0.013), but in fact, this trend was caused by cosmological dimming. After correcting for this effect, the trend is no longer measured (R < 0.1). We once again verify this result, as no correlation could be found here between the corrected mean surface brightness and redshift.
The large size of our sample enables us to confirm the results shown in \citet{Chu+21} up to $z$ = 0.7, namely that BCGs were mainly formed before 0.7 (z = 1.8 in our previous study), and have properties that appear to have remained stable since then.

Following the work of \citet{graham1997MNRAS.287..221G,bai2014inside}, we demonstrate that the S\'ersic index varies as the logarithm of the effective radius: $n = (5.13 \pm 0.21)$log(R$_e)+(−0.29 \pm 0.26).$
\noindent 
while \citet{graham1997MNRAS.287..221G} find approximately n $\propto$ 3.22 log(R$_{e}$). Our relation is steeper than that found by these authors, which means the measured S\'ersic indices are more sensitive to small variations of the effective radius.

We also plot the Kormendy relation for BCGs, which is very well defined with our sample. Our relation, not corrected for cosmological dimming:

$\rm <\mu> = (3.34 \pm 0.05) log R_{e} + (18.65 \pm 0.07)$

\noindent agrees within $1\sigma$ with that given in \citet{bai2014inside, Chu+21} and within three sigma with \citet{Durret_2019}:

$\rm <\mu> = (3.50 \pm 0.18) log \rm R_{e} + (18.01 \pm 0.23)$ \citep[][]{bai2014inside}

$\rm <\mu> = (3.33 \pm 0.73) log R_{e} + C$ \citep[][]{Chu+21}

$\rm <\mu> = (2.64 \pm 0.35) log \rm R_{e} + (19.7 \pm 0.5)$ \citep[][]{Durret_2019}

The dependence with redshift is due to cosmological dimming, which moves down the relation to fainter surface brightnesses without affecting the sizes of the BCGs.
The slope measured is also steeper than that of \citet{bai2014inside} measured for non BCG early type galaxies: 
$\rm <\mu> = (2.63 \pm 0.28) log \rm R_{e} + C$. 


Following the work of \citet{donahue2015,west2017ten, Durret_2019,  De_Propris_2020, Chu+21}, we show that the major axis of the BCG tends to align with that of the host cluster. Indeed, we find that at least $44 \pm 2\%$ of BCGs are aligned within 30 degrees with their host clusters. By only considering the best measured PAs (uncertainties smaller than 35 degrees), this percentage goes up to $57 \pm 4\%$. If BCGs had a random orientation, we would expect a uniform distribution and thus only $f_{\rm random} = 33\%$ of BCGs aligned within 30 deg with the major axis of their host clusters. We confirm that the measured alignment fractions are not due to random fluctuations due to the finite number of clusters studied. We also confirm results by \citet{Faltenbacher2009,10.1111/j.1365-2966.2010.16597.x} and \citet{Hao2011} who find stronger alignments for brighter and bigger galaxies.
In the hierarchical scenario of structure formation, matter and galaxies fall into the center of the cluster along cosmic filaments. This would create tidal interactions that can explain the observed alignment of BCGs with their host clusters. Thus, contrary to other cluster members, as was shown by \citet{west2017ten} who show that non BCGs members of a cluster have a random orientation in the cluster, the BCG properties are linked to the cluster.

This study shows, with increased statistics, evidence for an early formation of the brightest central galaxies in clusters. Most of their matter content was already in place by $z = 0.7$, and we showed in \citet{Chu+21} that this conclusion can most likely be applied up to $z = 1.8$. New datasets in the infrared (JWST, Euclid) should enable us to confirm this result at higher redshifts with better statistics. In the present paper, we also estimate in a first approach the contribution of the ICL to such studies. 


\begin{acknowledgements} 
We are very grateful to Y. Jim\'enez-Teja for providing us with her ICL images. We also thank A. Ellien for discussions on the ICL. We thank the referee for her/his constructive comments and suggestions.
  F.D. acknowledges continuous support from CNES since 2002. 
  I.M. acknowledges financial support from the State Agency for Research of the Spanish MCIN through the "Center of Excellence Severo Ochoa" award to the Instituto de Astrofísica de Andalucía (SEV-2017-0709), and through the program PID2019-106027GB-C41.
  Based on observations obtained with MegaPrime/MegaCam, a joint project of CFHT and CEA/DAPNIA, at the Canada-France-Hawaii Telescope (CFHT) which is operated by the National Research Council (NRC) of Canada, the Institut National des Sciences de l’Univers of the Centre National de la Recherche Scientifique (CNRS) of France, and the University of Hawaii. This work is based in part on data products produced at Terapix and the Canadian Astronomy Data Centre as part of the Canada-France-Hawaii Telescope Legacy Survey, a collaborative project of NRC and CNRS.

\end{acknowledgements}

\bibliographystyle{aa}
\bibliography{aa.bib}




\end{document}